\documentclass[twocolumn,showpacs,prb,aps,superscriptaddress]{revtex4}
\usepackage{amsmath}
\usepackage{graphicx}
\usepackage{dcolumn}

\newcommand{\be}{\begin{equation}}
\newcommand{\ee}{\end{equation}}
\newcommand{\bea}{\begin{eqnarray}}
\newcommand{\eea}{\end{eqnarray}}
\newcommand{\bse}{\begin{subequations}}
\newcommand{\ese}{\end{subequations}}

\begin{document}
\title{Physical properties of EuPd$_2$As$_2$ single crystals}
\author{V. K. Anand}
\altaffiliation{vivekkranand@gmail.com}
\affiliation {Ames Laboratory and Department of Physics and Astronomy, Iowa State University, Ames, Iowa 50011, USA}
\affiliation{Helmholtz-Zentrum Berlin f\"{u}r Materialien und Energie, Hahn-Meitner Platz 1, D-14109 Berlin, Germany}
\author{D. C. Johnston}
\altaffiliation{johnston@ameslab.gov}
\affiliation {Ames Laboratory and Department of Physics and Astronomy, Iowa State University, Ames, Iowa 50011, USA}

\date{\today}

\begin{abstract}
The physical properties of self-flux grown ${\rm EuPd_2As_2}$ single crystals have been investigated by magnetization $M$, magnetic susceptibility $\chi$, specific heat $C_{\rm p}$, and electrical resistivity $\rho$ measurements versus temperature $T$ and magnetic field $H$. The crystal structure was determined by powder x-ray diffraction measurements, which confirmed the ${\rm ThCr_2Si_2}$-type body-centered tetragonal structure (space group $I4/mmm$) reported previously.  The  $\rho(T)$ data indicate that state of ${\rm EuPd_2As_2}$ is  metallic.  Long-range antiferromagnetic (AFM) ordering is apparent from the $\chi(T)$, $C_{\rm p}(T)$, and $\rho (T)$ measurements. For $H \parallel c$ the $\chi(T)$ indicates two transitions at $T_{\rm N1}=11.0$~K and $T_{\rm N2}=5.5$~K, whereas for $H \perp c$ only one transition is observed at $T_{\rm N1}=11.0$~K\@. Between $T_{\rm N1}$ and~$T_{\rm N2}$ the anisotropic $\chi(T)$ data suggest a planar noncollinear AFM structure, whereas at $T < T_{\rm N2}$ the $\chi(T)$ and $M(H,T)$ data suggest a spin reorientation transition in which equal numbers of spins cant in opposite directions out of the $ab$~plane.  We estimate the critical field at 2~K at which all Eu moments become aligned with the field to be about 22~T\@.  The magnetic entropy at 25~K estimated from the $C_{\rm p}(T)$ measurements is about 11\% smaller than expected, possibly due to an inaccuracy in the lattice heat capacity contribution.  An upturn in $\rho$ at $T < T_{\rm N1}$ suggests superzone energy gap formation below $T_{\rm N1}$. This behavior of $\rho(T < T_{\rm N1})$ is not sensitive to applied magnetic fields up to $H = 12$~T\@.
\end{abstract}


\maketitle

\section{\label{Intro} INTRODUCTION}

The observation of high-temperature superconductivity upon suppression of magnetic long-range antiferromagnetic (AFM) spin density wave (SDW) ordering in iron arsenides such as K-doped $A$Fe$_2$As$_2$ ($A$ = Ba, Ca, Sr) with the body-centered-tetragonal ${\rm ThCr_2Si_2}$-type structure (space group $I4/mmm$) stimulated great interest in these materials. \cite{Rotter2008a, Chen2008a, Sasmal2008, Sefat2008, Torikachvili2008, Ishida2009, Alireza2009, Johnston2010, Canfield2010, Mandrus2010, Stewart2011} The introduction of local moments at the $A$ sites leads to coexistence of both intinerant and localized magnetic moments in the magnetically ordered state.  EuFe$_2$As$_2$ is a very nice example of this. While the itinerant carriers undergo an SDW transition at 190~K with the ordered moments concentrated near the Fe$^{+2}$ cation sites, the localized Eu$^{+2}$ moments with spin $S = 7/2$ order antiferromagnetically below 19~K. \cite{Ren2008, Jiang2009c, Xiao2009} The Eu sublattice has an A-type AFM structure where the ordered Eu$^{+2}$ moments in the $ab$~plane are aligned ferromagnetically and are aligned antiferromagnetically along the $c$~axis. \cite{Xiao2009} Similar to (Ba,Ca,Sr)Fe$_2$As$_2$, EuFe$_2$As$_2$ also exhibits superconductivity after the complete suppression of the itinerant SDW transition with $T_c$ as high as 33~K for Eu$_{0.5}$K$_{0.5}$Fe$_2$As$_2$. \cite{Jeevan2008, Ren2009a, Miclea2009, Jiang2009b, Jeevan2011, Anupam2011, Anupam2012} In addition, the presence of Eu moments provides an opportunity to explore the interplay and coexistence of long-range AFM order of the Eu spins and superconductivity in EuFe$_2$As$_2$ under pressure. \cite{Kurita2011}

Other compounds with Eu occupying the $A$~site of the ${\rm ThCr_2Si_2}$-type structure have been studied.  ${\rm EuFe_2P_2}$ orders ferromagnetically at $T = 30$~K with the Eu$^{+2}$ ordered moments canted at an angle of 17$^\circ$ from the $c$~axis and presents a dense Kondo behavior. \cite{Feng2010,Ryan2012} ${\rm EuCo_2P_2}$ has an AFM structure below $T_{\rm N}=66.5$~K with the Eu$^{+2}$ ordered moments aligned ferromagnetically in the $ab$ plane, forming an incommensurate AFM spiral structure along the $c$~axis. \cite{Reehuis1992} In ${\rm EuCo_2P_2}$, the magnetic ordering of Eu$^{+2}$ is suppressed under pressure with a simultaneous magnetic ordering of itinerant carriers at $T_{\rm N}=260$~K with the ordered moments centered on the Co sites at a critical pressure p$_c$ = 3.1~GPa, where a pressure-induced isostructural phase transition from a tetragonal (T) phase to collapsed tetragonal (cT) phase also occurs. \cite{Huhnt1997,Chefki1998} ${\rm EuCo_2As_2}$ is reported to exhibit AFM ordering below 39~K for which an A-type AFM structure is proposed. \cite{Ballinger2012} ${\rm EuCo_2As_2}$ also exhibits a pressure-induced isostructural phase transition from the T phase to cT phase at 4.7~GPa. \cite{Bishop2010} ${\rm EuCu_2As_2}$ is found to order antiferromagnetically below $T_{\rm N} = 15$~K. \cite{Sengupta2005}  A strong increase is observed in the ordering temperature of ${\rm EuCu_2As_2}$ from 15~K at ambient pressure to 49~K at 10.7~GPa with a possible crossover from AFM structure to a ferromagnetic (FM) structure above 7~T. \cite{Sengupta2012} Our investigations on single crystal ${\rm EuCu_2As_2}$ with the ${\rm ThCr_2Si_2}$-type structure and ${\rm EuCu_2Sb_2}$ with the different primitive tetragonal ${\rm CaBe_2Ge_2}$-type structure revealed AFM ordering of the Eu$^{+2}$ moments in both compounds below $T_{\rm N} = 17.5$~K and 5.1~K, respectively. \cite{Anand2014a}  While the $\chi(T)$ data suggest that ${\rm EuCu_2Sb_2}$ has an A-type AFM structure, the AFM structure of ${\rm EuCu_2As_2}$ is unclear as yet. \cite{Anand2014a}

We previously investigated the physical properties of ${\rm EuPd_2Sb_2}$ with the primitive tetragonal ${\rm CaBe_2Ge_2}$-type structure, which is closely related to the ${\rm ThCr_2Si_2}$-type structure. \cite{Das2010}  This compound shows AFM ordering of the Eu spins at $T_{\rm N1} = 6.0$~K with another AFM transition at $T_{\rm N2} = 4.5$~K that may be a spin-reorientation transition.  From single-crystal $\chi(T)$ measurements, the compound appears to have a noncollinear AFM structure.  We also studied $A{\rm Pd_2As_2}$ ($A$ = Ca, Sr, and Ba) with the ${\rm ThCr_2Si_2}$-type structure and discovered bulk superconductivity in ${\rm CaPd_2As_2}$ and ${\rm SrPd_2As_2}$ below $T_{\rm c} = 1.27$ and 0.92~K, respectively. \cite{Anand2013a}

${\rm EuPd_2As_2}$ also crystallizes in the ${\rm ThCr_2Si_2}$-type structure. \cite{Hofmann1985} A preliminary investigation of the magnetic properties of ${\rm EuPd_2As_2}$ using $\chi(T)$ and M\"ossbauer measurements revealed AFM ordering of the Eu moments in a polycrystalline sample below $T_{\rm N} = 11$~K. \cite{Raffius1993}  We have grown single crystals of ${\rm EuPd_2As_2}$ by the self-flux method and present herein their physical properties obtained from magnetic susceptibility $\chi$, isothermal magnetization $M$, heat capacity $C_{\rm p}$ and electrical resistivity $\rho$ measurements as a function of temperature $T$ and magnetic field $H$\@.

We confirm the presence of Eu$^{+2}$ magnetic moments with $S = 7/2$ and spectroscopic splitting factor $g = 2$ and AFM ordering of these spins below $T_{\rm N} = 11$~K as found in Ref.~\onlinecite{Raffius1993}.  We report an additional transition at 5.5~K that is likely due to an AFM spin reorientation transition.  The $\chi(T)$ measured in low~$H$ exhibits two transitions at $T_{\rm N1}=11.0$~K and $T_{\rm N2}=5.5$~K for $H \parallel c$, and one transition at $T_{\rm N1}=11.0$~K for $H \perp c$. The $M(H)$ at~2~K up to $H=14$~T shows a weak upward curvature, consistent with an AFM structure. The $C_{\rm p}(T)$ data show a sharp $\lambda$-type anomaly at $T_{\rm N1}$, whereas the anomaly at $T_{\rm N2}$ is weaker. The $\rho (T)$ data demonstrate that ${\rm EuPd_2As_2}$ is metallic and the data show anomalies at both $T_{\rm N1}$ and $T_{\rm N2}$. The $\rho (T)$ exhibits a sharp upturn below $T_{\rm N1}$ possibly due to the formation of a superzone energy gap over part of the Brillouin zone at $T_{\rm N1}$.  No change in the upturn in $\rho (T)$ is evident under applied magnetic fields up to $H =12$~T\@. In the paramagnetic state above 16~K the $C_{\rm p}(T)$ data are well represented by the Debye model of lattice heat capacity and the $\rho(T)$ data by the Bloch-Gr\"{u}neisen model for the contribution to $\rho(T)$ from electron-phonon scattering.

\section{\label{ExpDetails} EXPERIMENTAL DETAILS}

Single crystals of ${\rm EuPd_2As_2}$ were grown by the high-temperature solution growth method using self-flux. High-purity Eu (Ames Laboratory) and prereacted PdAs [Pd (99.998\%) and As (99.99999\%), Alfa Aesar] taken in a 1:5 molar ratio were placed in an alumina crucible and sealed inside an evacuated quartz tube. The sealed sample was heated to 1100~$^\circ$C at a rate of 60~$^\circ$C/h and held there for 15~h, followed by  cooling at a rate of 2.5~$^\circ$C/h to 800~$^\circ$C at which point the flux was decanted with a centrifuge, yielding shiny plate-like crystals of typical size $2 \times 1.5 \times 0.4$~mm$^3$.

The chemical composition and quality of the crystals were checked using a JEOL scanning electron microscope (SEM) equipped with an energy dispersive x-ray (EDX) analyzer. The SEM images indicated from the uniformity of the (001) plane faces that the crystals contain only a single phase.  The  EDX composition analysis confirmed the desired stoichiometry of the crystals with Eu:Pd:As in a 1\,:\,2\,:\,2 molar ratio. The crystal structure was determined by powder x-ray diffraction (XRD) using Cu~K$_\alpha$ radiation on a Rigaku Geigerflex x-ray Diffractometer. The XRD data were refined by Rietveld refinement using the {\tt FullProf} software package. \cite{Rodriguez1993}

The $\chi(T)\equiv M(T)/H$ and $M(H)$ isotherms were measured using a Quantum Design, Inc., superconducting quantum interference device (SQUID) magnetic properties measurement system (MPMS). $M(H)$ isotherms at high magnetic field were measured using the vibrating sample magnetometer (VSM) option of a Quantum Design, Inc., physical properties measurement system (PPMS). The sample holder contributions to the measured magnetic moments were subtracted to obtain the sample contributions. The magnetic properties are expressed exclusively in Gaussian cgs units, where the Tesla (T) is a common unit of convenience for the magnetic field~$H$ defined as 1~T = $10^4$~Oe.  The $C_{\rm p}(T)$ was measured by a relaxation method using the heat capacity option of the PPMS\@. The $\rho(T)$ was measured by the standard four-probe ac technique using the ac transport option of the PPMS\@.

\section{\label{EuPd2As2} Results and Discussion}

\subsection{\label{Sec:EuPd2AS2_XRD} Crystallography}

\begin{figure}
\includegraphics[width=3in, keepaspectratio]{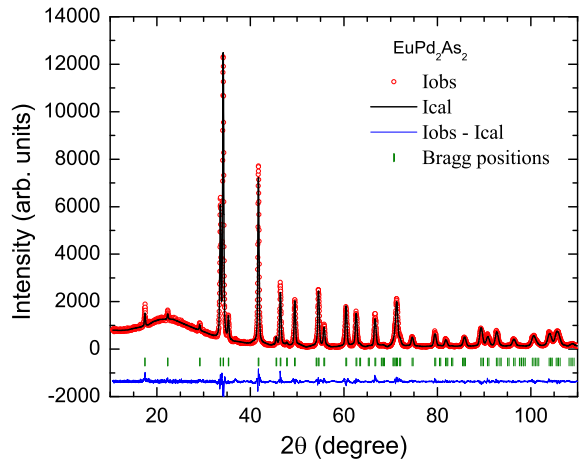}
\caption {(Color online) Powder x-ray diffraction pattern of ${\rm EuPd_2As_2}$ recorded at room temperature. The solid line through the experimental points is the Rietveld refinement profile calculated for the ${\rm ThCr_2Si_2}$-type body-centered tetragonal structure (space group $I4/mmm$). The short vertical bars mark the Bragg peak positions. The lowermost curve represents the difference between the experimental and calculated intensities.}
\label{fig:XRD}
\end{figure}

Powder x-ray diffraction data collected on crushed ${\rm EuPd_2As_2}$ single crystals at room temperature are shown in Fig.~\ref{fig:XRD} together with the Rietveld refinement profile.  The refinement confirmed the ${\rm ThCr_2Si_2}$-type body-centered tetragonal structure (space group $I4/mmm$) of ${\rm EuPd_2As_2}$ and showed no impurity peaks.  While refining, the thermal parameters $B \equiv 0$ and the fractional occupancies were fixed to unity.  Small variations in $B$ ($\lesssim 0.5$~\AA$^2$) and in the occupancies of atomic positions ($\lesssim10$\%) had no noticeable effect on the quality of fit or on the refined lattice parameters and $z_{\rm As}$. The crystallographic parameters are listed in Table~\ref{tab:XRD1}.  The lattice parameters are in good agreement with the literature values. \cite{Hofmann1985} The interlayer As--As distance $d_{\rm As-As} = (1-2z_{\rm As})c = 2.498$~\AA\ and $c/a = 2.3488(3)$ are close to values typical for collapsed tetragonal compounds as discussed in Ref.~\onlinecite{Anand2012a}, indicating that ${\rm EuPd_2As_2}$ has a collapsed tetragonal structure.

\begin{table}
\caption{Crystallographic and Rietveld refinement parameters obtained from powder XRD data of crushed ${\rm EuPd_2As_2}$ crystals with the body-centered tetragonal ${\rm ThCr_2Si_2}$-type structure with space group  $I4/mmm$. The atomic coordinates of Eu, Pd and As atoms are (0,0,0), (0,1/2,1/4) and (0,0,$z_{\rm As}$), respectively.}
\label{tab:XRD1}
\begin{ruledtabular}
\begin{tabular}{llll}
\underline{Lattice parameters}\\
\hspace{0.8cm} $a$ (\AA)            		&  4.3298(2)  \\	
\hspace{0.8cm} $c$ (\AA)          			&  10.1700(3) \\
\hspace{0.8cm} $V_{\rm cell}$  (\AA$^3$) 	&  190.66(1)  \\
As $c$ axis coordinate $z_{\rm As}$   &  0.3772(2) \\
\\
\underline{Refinement quality} \\
\hspace{0.8cm} $\chi^2$   & 3.09\\	
\hspace{0.8cm} $R_{\rm p}$ (\%)  & 5.33\\
\hspace{0.8cm} $R_{\rm wp}$ (\%) & 7.46\\
\end{tabular}
\end{ruledtabular}
\end{table}

\subsection{\label{Sec:EuPd2AS2_ChiMH} Magnetization and Magnetic Susceptibility}

\subsubsection{High-Temperature Paramagnetic Susceptibility}

\begin{figure}
\includegraphics[width=3in]{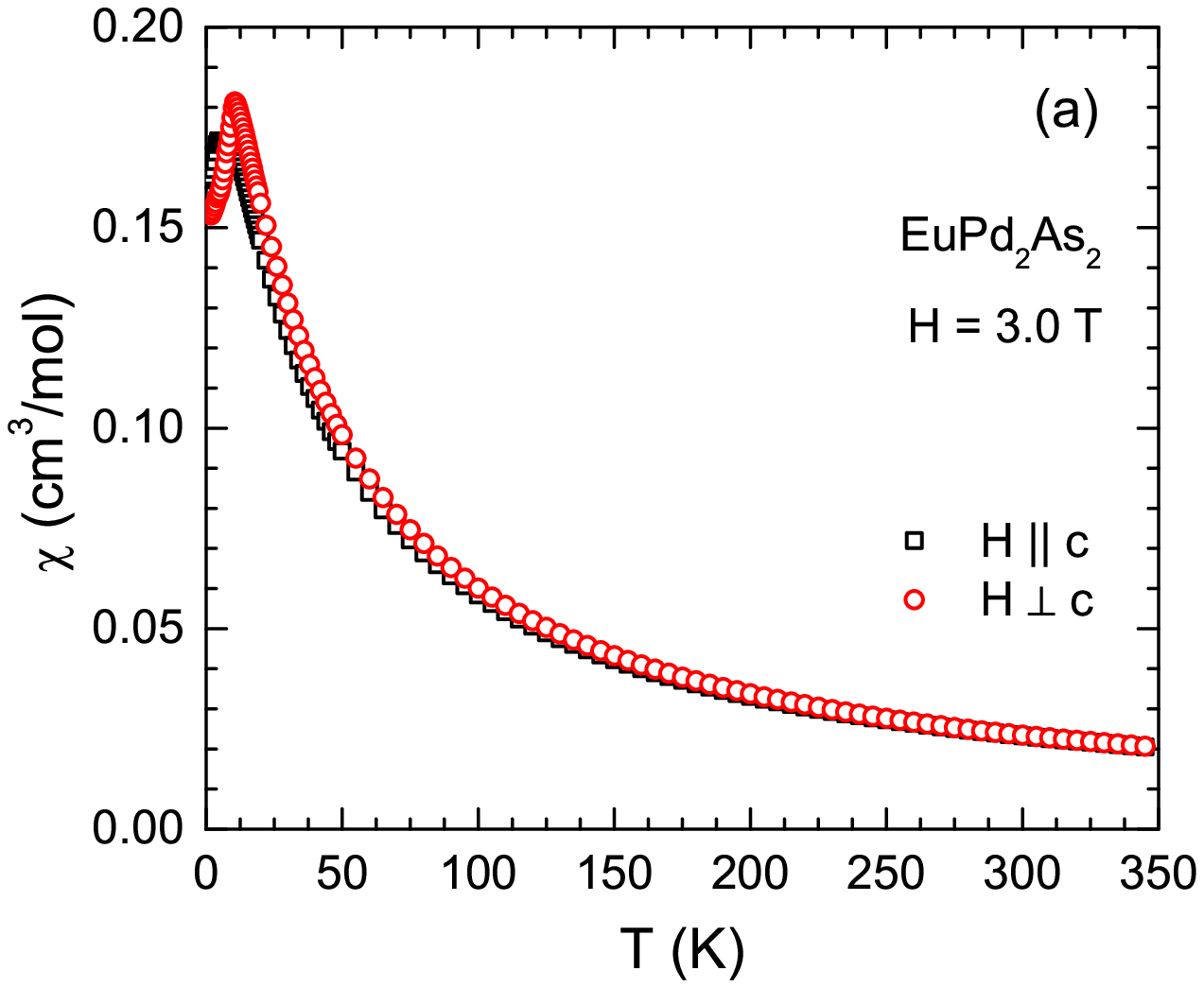}
\includegraphics[width=3in]{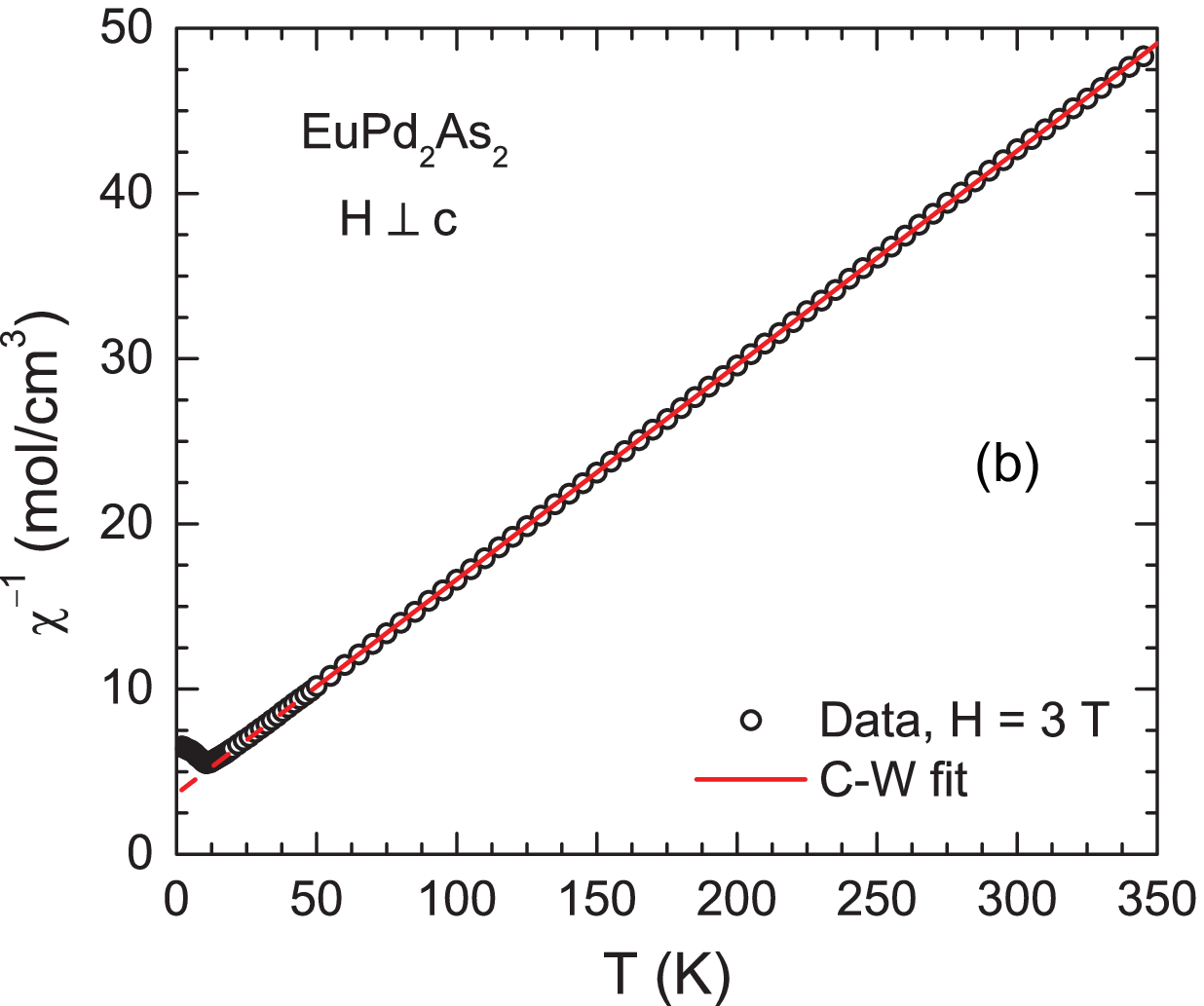}
\includegraphics[width=3in]{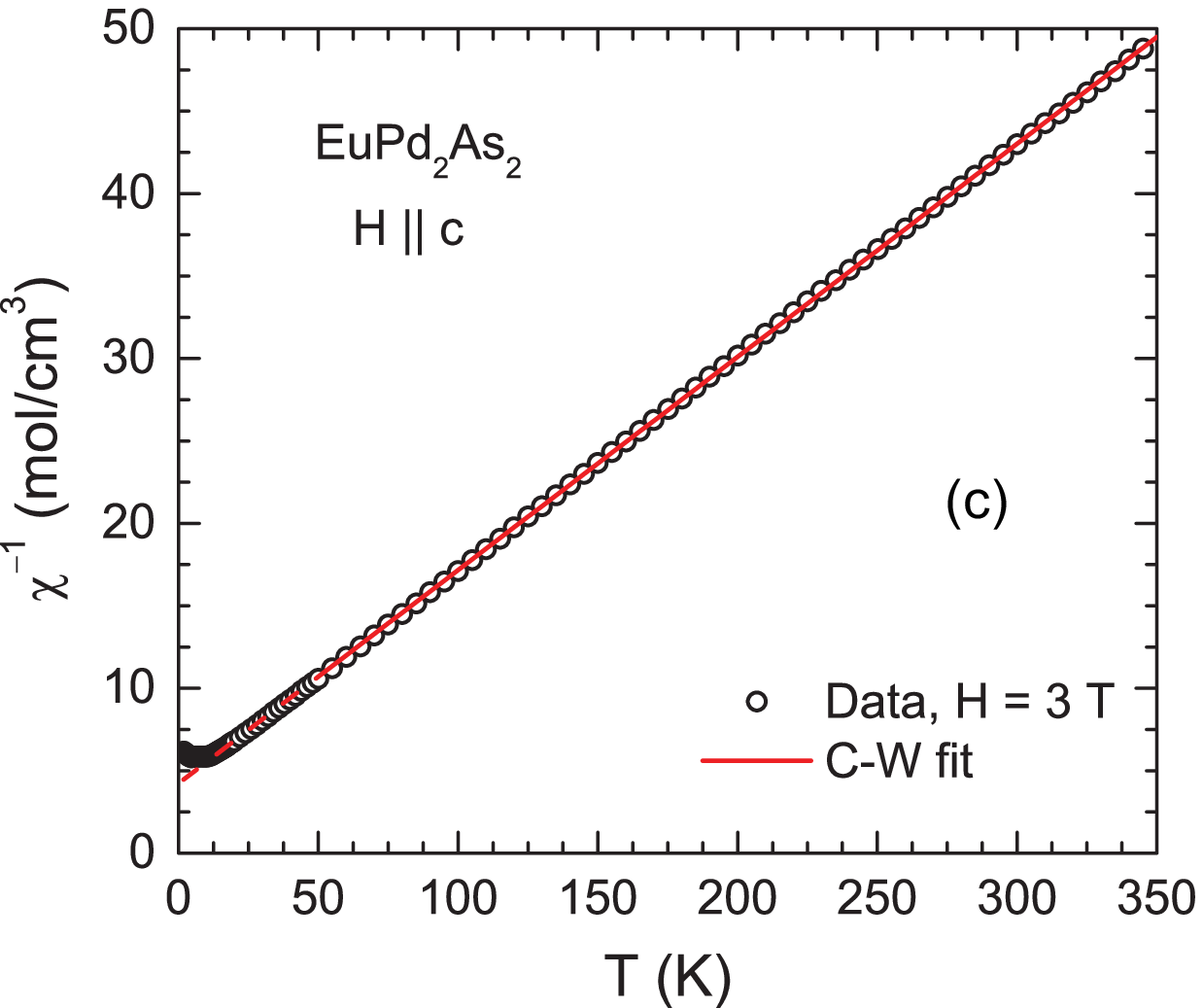}
\caption{(Color online) (a) Zero-field-cooled magnetic susceptibility $\chi$ of a ${\rm EuPd_2As_2}$ single crystal as a function of temperature $T$ in the temperature range 1.8--350~K measured in a magnetic field $H=3.0$~T applied in the $ab$ plane ($\chi_{ab}, H \perp c$) and along the $c$ axis ($\chi_c, H \parallel c$). (b) $\chi^{-1}(T)$ for $H \perp c$, and (c) $\chi^{-1}(T)$ for $H \parallel c$. The red solid straight lines in (b) and (c) are fits of the respective $\chi^{-1}(T)$ data by the Curie-Weiss (CW) law in the temperture range 50~K~$\leq T \leq$~350~K\@. The red dashed straight lines are extrapolations of the CW fits to lower temperatures.}
\label{fig:MT_EuPd2As2_HighT}
\end{figure}

\begin{table}
\caption{Magnetic ordering temperatures $T_{\rm N1}$ and $T_{\rm N2}$ measured from the low-field susceptibility data in Fig.~\ref{fig:MT_EuPd2As2LowT}(a) and the Curie constants~$C$, the Weiss temperatures~$\theta_{\rm p}$ and effective moments per Eu $\mu_{\rm eff} = \sqrt{8C}$  obtained from  Curie-Weiss fits to the high-temperature $\chi(T)$ data for ${\rm EuPd_2As_2}$ in Figs.~\ref{fig:MT_EuPd2As2_HighT}(b) and~\ref{fig:MT_EuPd2As2_HighT}(c).}
\label{tab:CW}
\begin{ruledtabular}
\begin{tabular}{cccccc}
Field   & $T_{\rm N1}$  & $T_{\rm N2}$ & $C$  &  $\theta_{\rm p}$   &	$\mu_{\rm eff}$ \\
direction & (K) & (K) & (cm$^3$\,K/mol) & (K) & ($\mu_{\rm B}$/Eu) \\
\hline
$H \parallel c$  & 11.0 & 5.5 & 7.73(3)  & $-$32.8(9) &  7.86(2)  \\	
$H \perp c$      & 11.0 &     & 7.71(2)  & $-$28.1(3) & 7.85(1)   \\				
\end{tabular}
\end{ruledtabular}
\end{table}

The $\chi(T)\equiv M(T)/H$ data measured for ${\rm EuPd_2As_2}$ in $H = 3$~T for $H\parallel c$ and $H\perp c$ up to 350~K are shown in Fig.~\ref{fig:MT_EuPd2As2_HighT}(a).  The data are nearly isotropic on the scale of the figure.  An AFM transition is seen at low temperatures $\leq10$~K [see also Fig.~\ref{fig:MT_EuPd2As2LowT}(a) below and Table~\ref{tab:CW}].  The data in the paramagnetic state follow the Curie-Weiss law $\chi(T) = C/(T-\theta_{\rm p})$, where $C$ is Curie constant and $\theta_{\rm p}$ is the Weiss temperature.  The $1/\chi$ versus~$T$ data for $H\perp c$ and $H\parallel c$ are shown in Figs.~\ref{fig:MT_EuPd2As2_HighT}(b) and~\ref{fig:MT_EuPd2As2_HighT}(c), respectively.  Linear fits of these two sets of data by the inverse Curie-Weiss law for 50~K~$\leq T \leq$~350~K are shown as straight lines in the respective figures. The fits yield $C = 7.71(2)$~cm$^3$\,K/mol and $\theta_{\rm p}^{ab}= -28.1(3)$~K for $H \perp c$ and $C = 7.73(2)$~cm$^3$\,K/mol and $\theta_{\rm p}^c = -32.8(9)$~K for $H \parallel c$. The negative $\theta_{\rm p}$ values indicate that the dominant magnetic interactions in ${\rm EuPd_2As_2}$ are AFM\@. The Curie constant calculated for Eu$^{+2}$ cations with $S=7/2$ and spectroscopic splitting factor $g=2$ is $C^{\rm calc} = 7.88$~cm$^3$\,K/mol\,Eu, which is very close to the measured values.  We conclude that the Eu in ${\rm EuPd_2As_2}$ is in the +2 oxidation state with $S = 7/2$ and $g=2$.  The parameters obtained from the Curie-Weiss fits of the $\chi^{-1}(T)$ data are summarized in Table~\ref{tab:CW}.

\subsubsection{Low-Temperature Magnetic Susceptibility}

\begin{figure}
\includegraphics[width=2.75in]{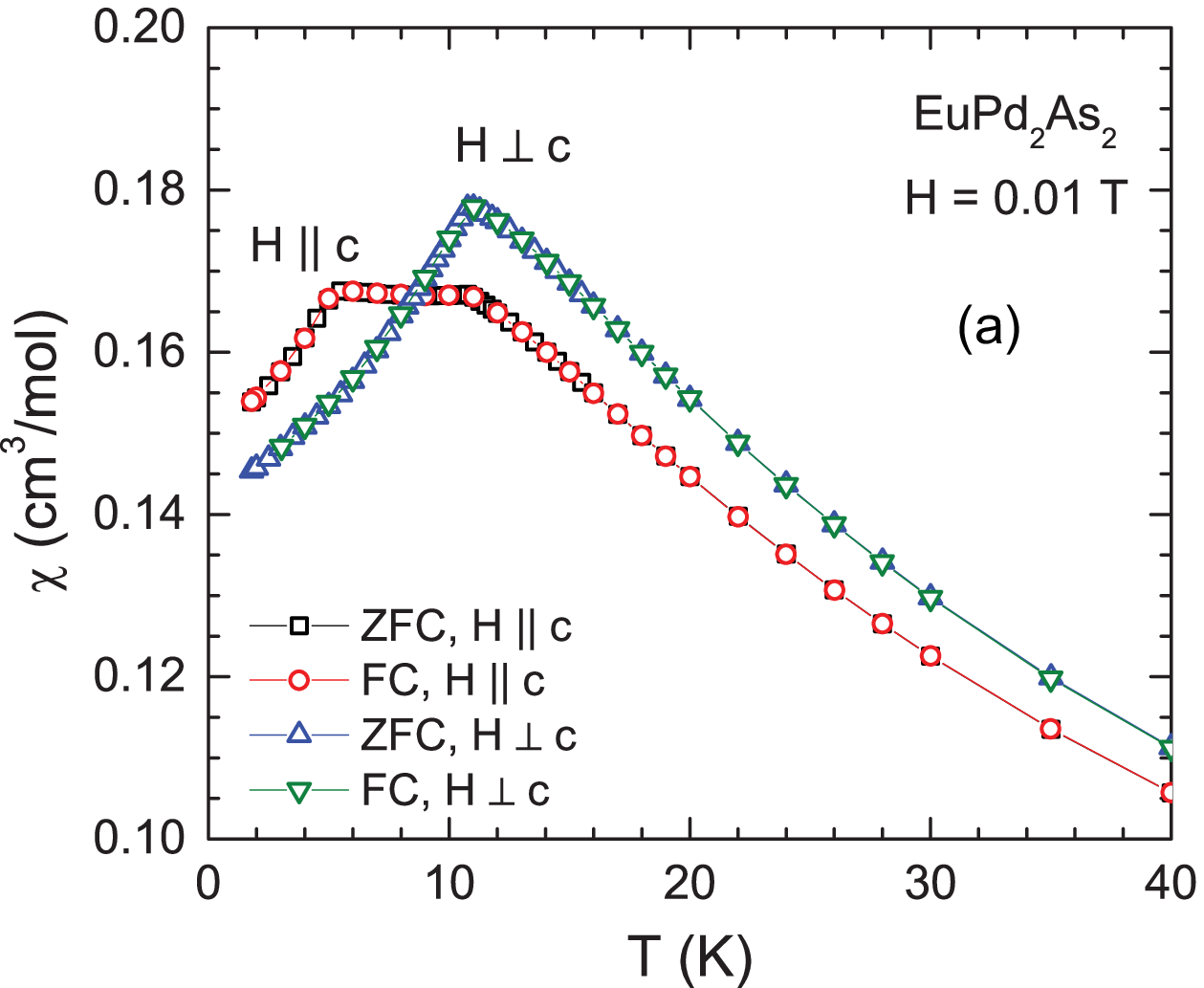}
\includegraphics[width=2.75in]{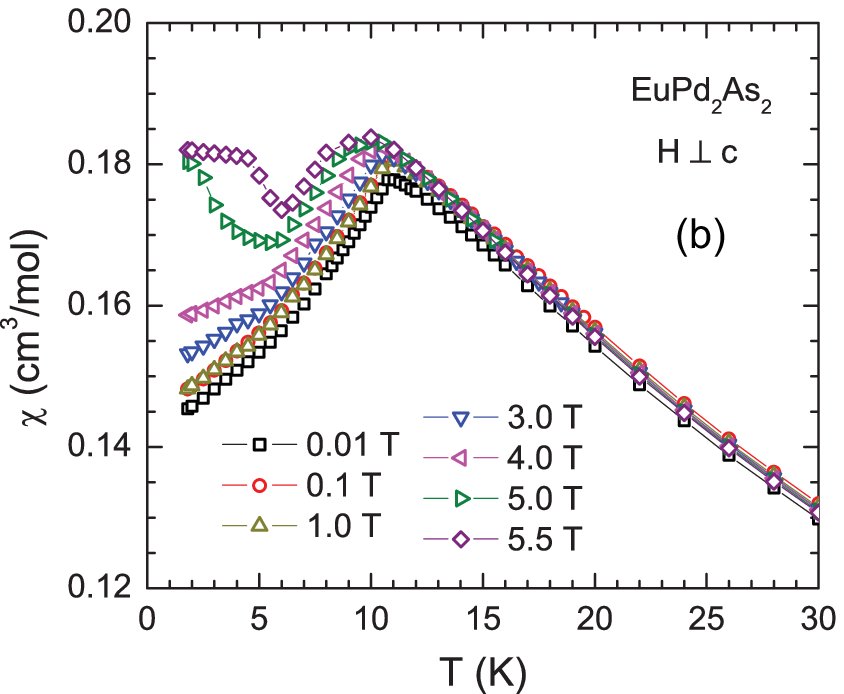}
\includegraphics[width=2.75in]{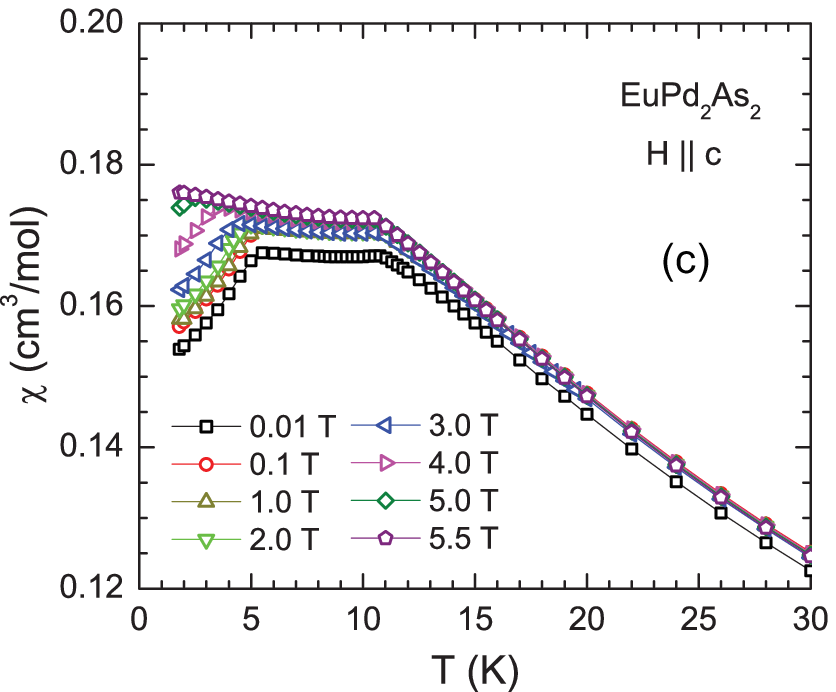}\vspace{-0.1in}
\caption{(Color online) (a) Zero-field-cooled (ZFC) and field-cooled (FC) magnetic susceptibility $\chi$ of a ${\rm EuPd_2As_2}$ single crystal versus temperature $T < 40$~K measured in $H= 0.01$~T applied along the $c$ axis ($\chi_c, H \parallel c$) and in the $ab$ plane ($\chi_{ab}, H \perp  c$).  (b) ZFC $\chi$ versus $T$ in the temperature range 1.8--30~K measured in different magnetic field $H$ applied in the $ab$ plane ($\chi_{ab}, H \perp  c$) and (c) along the $c$ axis ($\chi_c, H \parallel c$).  Note the expanded vertical scales in all three panels.}
\label{fig:MT_EuPd2As2LowT}
\end{figure}

The zero-field-cooled (ZFC) and field-cooled (FC) $\chi  \equiv M/H$ of an ${\rm EuPd_2As_2}$ single crystal versus~$T$ measured at $H = 0.01$~T aligned along the $c$ axis ($\chi_c,\ H \parallel c$) and in the $ab$ plane ($\chi_{ab},\  H \perp c$) are shown in Fig.~\ref{fig:MT_EuPd2As2LowT}(a).  No thermal hysteresis between the ZFC and FC data is observed.  For $H \parallel c$, well-defined cusps are seen in the low-field $\chi(T)$ data at 11.0~K and 5.5~K whereas for $H \perp c$ only one cusp is observed at 11.0~K\@. Furthermore, the $\chi(T)$ data measured at different $H$ in Figs.~\ref{fig:MT_EuPd2As2LowT}(b) and~\ref{fig:MT_EuPd2As2LowT}(c) show that an increase in $H$ shifts these anomalies towards lower temperatures suggesting that the $\chi(T)$ cusps are due to AFM ordering.  We infer that two zero-field AFM transitions occur at $T_{\rm N1} = 11.0$~K and $T_{\rm N2}=5.5$~K\@.  While $T_{\rm N1}$ is due to a transition from paramagnetic phase to an AFM phase, $T_{\rm N2}$ may be associated with an AFM spin reorientation transition.

For $H \perp c$ only a barely detectable change in slope is observed near $T_{\rm N2} = 5.5$~K in the $\chi_{ab}(T)$ data at $H = 0.01$~T in Fig.~\ref{fig:MT_EuPd2As2LowT}(a).  However, as shown in Fig.~\ref{fig:MT_EuPd2As2LowT}(b), as $H$ increases the slope change becomes clearly observable, and at $H = 5.5$~T a well-defined anomaly can be seen in $\chi(T)$ at the same temperature 5.5~K\@.   Thus $T_{\rm N2}$ shows no detectable field dependence within our field range for $H \perp c$.  In contrast, $T_{\rm N1}$ for $H \perp c$ decreases significantly from 11.0~K at $H = 0.01$~T to $\approx 9.0$~K at $H = 5.0$~T\@.

From the $\chi_c(T)$ data in Fig.~\ref{fig:MT_EuPd2As2LowT}(c) with $H \parallel c$, both $T_{\rm N1}$ and $T_{\rm N2}$ decrease with increasing $H$\@. The $T_{\rm N1}$ decreases from 11.0~K at $H = 0.01$~T to 10.5~K at $H = 5.0$~T and $T_{\rm N2}$ decreases from 5.5~K at $H = 0.01$~T to $2.2$~K at $H = 5.0$~T\@.  Thus the change in $T_{\rm N2}$ with increasing $H$ is much larger than the change in $T_{\rm N2}$ discussed in the previous paragraph for $H \perp c$.

The low-field $\chi_c(T)$ data in Fig.~\ref{fig:MT_EuPd2As2LowT}(a) are temperature-independent between $T_{\rm N1}$ and~$T_{\rm N2}$, whereas the $\chi_{ab}(T)$ data decrease rapidly below  $T_{\rm N1}$.  Within the Weiss molecular field theory (MFT), this difference indicates that the AFM ordered moments lie in the tetragonal $ab$~plane. \cite{Johnston2011a}  The observation that $\chi_c<\chi_{ab}$ at $T>T_{\rm N1}$ and $\chi_c>\chi_{ab}$ at $T< 8.5$~K suggests the presence of a small anisotropy field parallel to the $ab$~plane both above and below $T_{\rm N1}$.  On the other hand, for a collinear AFM structure with the ordered moments in the $ab$~plane and equal numbers of AFM domains with their collinear axes at 90$^\circ$ to each other, one expects $\chi_{ab}(T\to0)/\chi_{ab}(T_{\rm N1}) = 1/2$, which is not realized in the data which show $\chi_{ab}(T\to0)/\chi_{ab}(T_{\rm N1}) \approx 0.80$. This large deviation from expectation for collinear AFM ordering suggests that the AFM structure of ${\rm EuPd_2As_2}$ between $T_{\rm N1}$ and $T_{\rm N2}$ is a noncollinear planar helical or cycloidal structure with the ordered moments aligned in the $ab$~plane.  The turn angle in MFT is a two-valued function of $\chi_{ab}(T\to0)/\chi_{ab}(T_{\rm N1})$ if $1/2 < \chi_{ab}(T\to0)/\chi_{ab}(T_{\rm N1}) < 1$.  With the observed value $\chi_{ab}(T\to0)/\chi_{ab}(T_{\rm N1}) \approx 0.80$, one obtains a turn angle of either $\sim104^\circ$ or 139$^\circ$ along the helix/cycloid axis between ferromagnetically aligned planes perpendicular to this axis. \cite{Johnston2012}  Our measurements cannot distinguish between the helical and cycloidal types of noncollinear AFM $ab$~plane ordering.  In helical ordering, the spin rotation (helix) axis is along the $c$~axis, whereas for cycloidal ordering, the spin rotation (cycloid) axis is in the $ab$~plane.

The decrease in $\chi_c$ at $T < T_{\rm N2}$ in Fig.~\ref{fig:MT_EuPd2As2LowT}(a) suggests that the in-plane moments become canted towards the $c$ axis in such a way as to retain the overall AFM structure, such as in a sequence of canted-up/canted-down spins out of the $ab$~plane.  Furthermore, the ordered-state $M(H)$ data presented in the following section exhibit upward curvature for both $H \perp c$ and $H \parallel c$, consistent with this canted AFM structure below $T_{\rm N2}$.

\subsubsection{Magnetization versus Applied Magnetic Field Isotherms}

\begin{figure}
\includegraphics[width=3in]{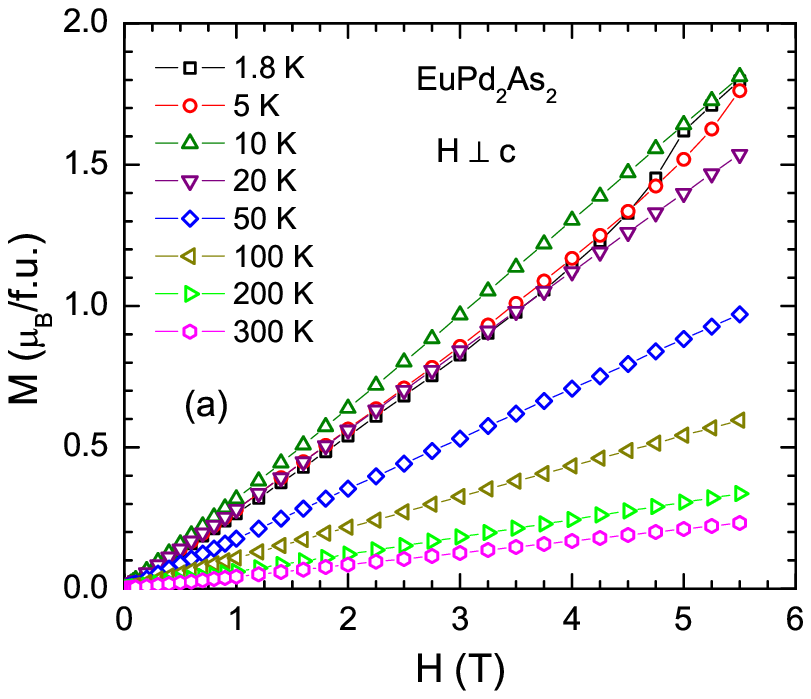}\vspace{0.1in}
\includegraphics[width=3in]{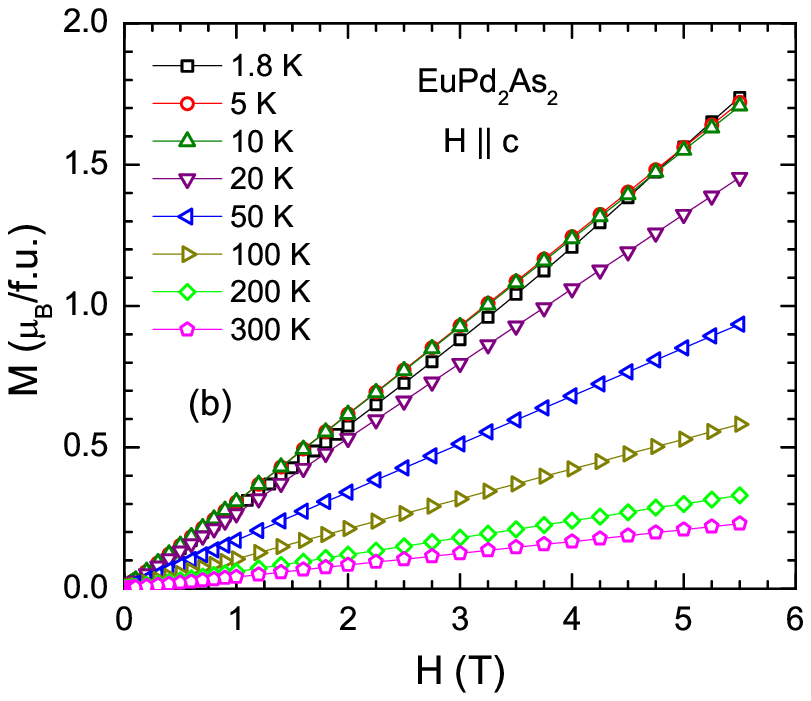}
\caption{(Color online) Magnetization $M$ versus applied magnetic field~$H$ isotherms of a ${\rm EuPd_2As_2}$  single crystal measured at the indicated temperatures for $H$ applied (a) in the $ab$ plane ($M_{ab}, H \perp  c$) and, (b) along the $c$ axis ($M_c, H \parallel c$).}
\label{fig:MH_EuPd2As2}
\end{figure}

\begin{figure}
\includegraphics[width=3in]{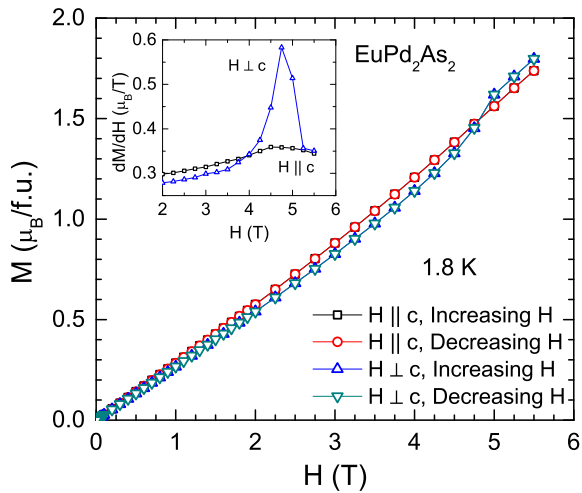}
\caption{(Color online) Isothermal magnetization $M$ of a ${\rm EuPd_2As_2}$  single crystal as a function of applied magnetic field $H$ measured at 1.8~K for $H$ applied in the $ab$ plane ($M_{ab}, H \perp  c$) and along the $c$ axis ($M_c, H \parallel c$). Inset: The field derivatives $dM_{ab}/dH$ and $dM_c/dH$ versus~$H$.}
\label{fig:MH_EuPd2As2_2K}
\end{figure}

Isothermal $M(H)$ data for an ${\rm EuPd_2As_2}$ crystal at eight temperatures between 1.8~and 300~K for $H$ applied both along the $c$ axis ($M_c, H \parallel c$) and in the $ab$ plane ($M_{ab}, H \perp  c$) are shown in Fig.~\ref{fig:MH_EuPd2As2} and data at 1.8~K for both increasing and decreasing $H$ are shown in Fig.~\ref{fig:MH_EuPd2As2_2K}, where $H \leq5.5$~T in both figures. The magnetization does not show saturation behavior up to $H=5.5$~T for either field direction.  It is seen from Fig.~\ref{fig:MH_EuPd2As2_2K} that at 1.8~K initially the $M$ exhibits almost a linear $H$ dependence for $H \leq 2.0$~T above which an upward curvature is seen for both $H \perp  c$ and $H \parallel c$ without any hysteresis between increasing and decreasing~$H$\@. The $M_{ab}$ resembles a weak $S$-shaped metamagnetic behavior. The derivative $dM/dH~{\rm versus}~H$ shown in the inset of Fig.~\ref{fig:MH_EuPd2As2_2K} clearly reflects this behavior, where a pronounced peak is observed at $H = 4.75$~T for $H \perp  c$. The weak change in slope for $H \parallel c$ is also evident from a broad peak near 4.5~T in $dM/dH~{\rm versus}~H$\@.  The observed magnetizations $M_{ab} = 1.80\,\mu_{\rm B}$/Eu and $M_{c} = 1.74\,\mu_{\rm B}$/Eu at $H=5.5$~T for $H \perp  c$ and $H \parallel c$, respectively, are much smaller than the theoretical value $M_{\rm sat} = 7\,\mu_{\rm B}$/Eu for $S = 7/2$ and $g=2$. Figure~\ref{fig:MH_EuPd2As2} shows that similar $M(H)$ behaviors are observed for $M_{ab}$ and $M_c$ at $T = 5$~K as at 1.8~K\@.  For  $T > T_{\rm N1}$ the $M$ is almost proportional to $H$ at fixed~$T$\@.

Within MFT, the critical field $H^{\rm c}$ of an AFM, which is the field at which $M$ reaches $M_{\rm sat}$ with increasing~$H$, is given by
\be
H^{\rm c} = \frac{M_{\rm sat}}{\chi(T_{\rm N})}.
\label{Eq:HcDef}
\ee
From Fig.~\ref{fig:MT_EuPd2As2LowT}(a), for $H \perp  c$ one has $\chi_{ab}(T_{\rm N})\approx 0.18~{\rm cm^3/mol} = 3.2\times10^{-5}~\mu_{\rm B}$/Oe\,Eu.  Then using $M_{\rm sat}=7\,\mu_{\rm B}$/Eu, Eq.~(\ref{Eq:HcDef}) gives the calculated critical field as
\be
H^{\rm c}_{ab} \approx 22~{\rm T}.
\label{Eq:HcCalc}
\ee
This value is a factor of four larger than our maximum measurement field of 5.5~T in Figs.~\ref{fig:MH_EuPd2As2} and~\ref{fig:MH_EuPd2As2_2K}.

\begin{figure}
\includegraphics[width=3in]{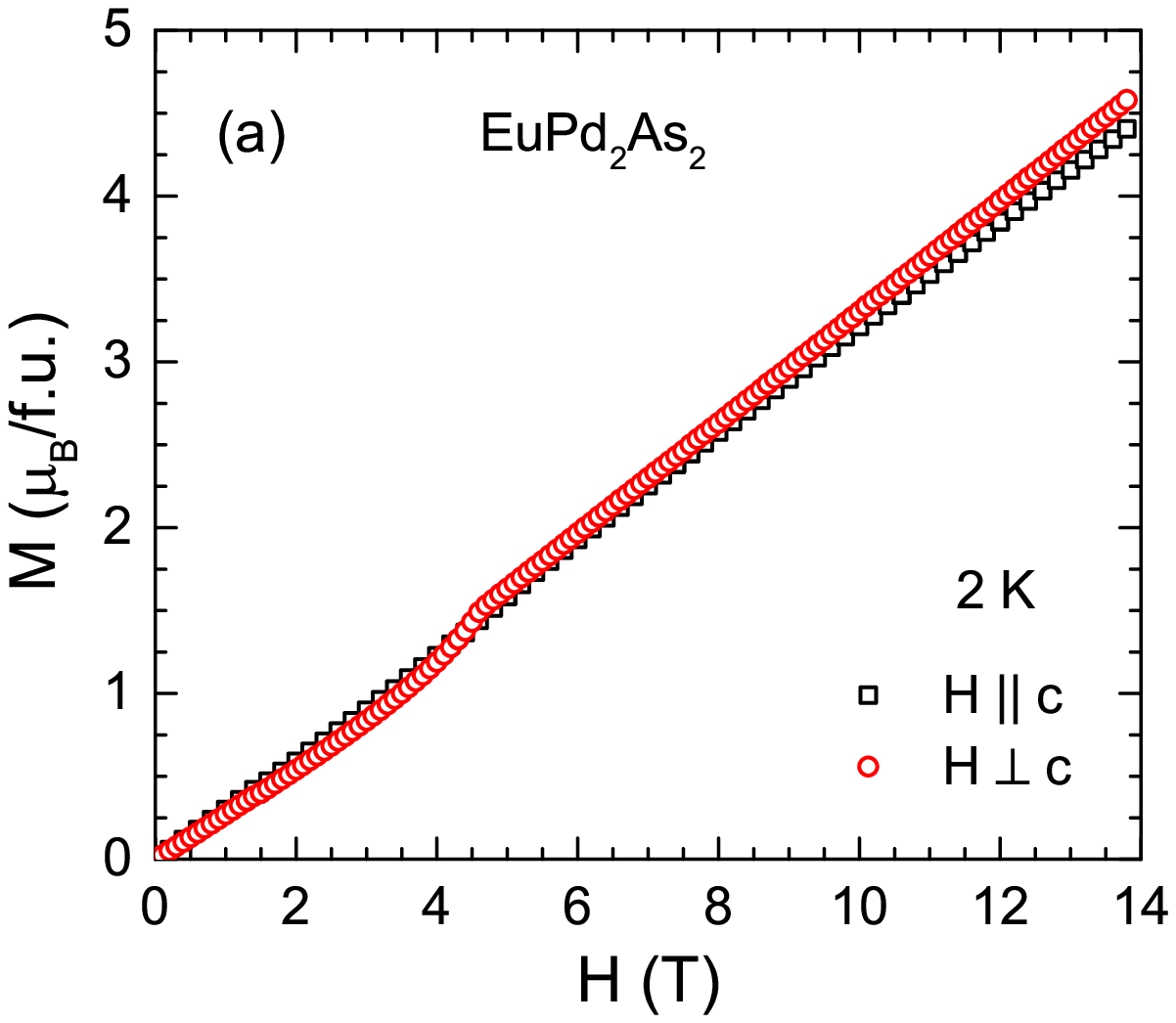}\vspace{0.1in}\vspace{-0.1in}
\includegraphics[width=3in]{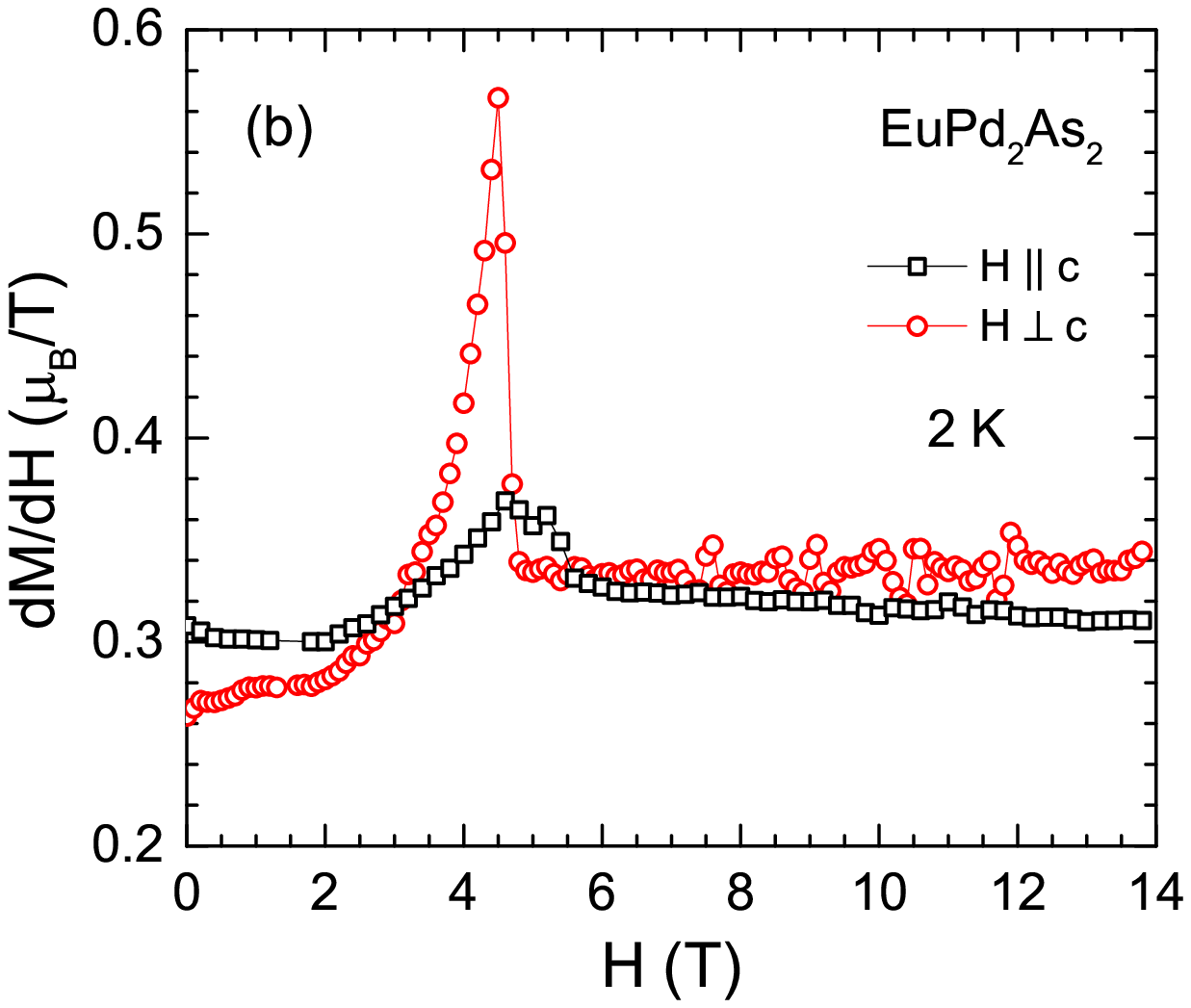}
\caption{(Color online) (a) Isothermal magnetization $M$ of a ${\rm EuPd_2As_2}$  single crystal as a function of applied magnetic field $H$ measured at 1.8~K for $H$ applied in the $ab$ plane ($M_{ab}, H \perp  c$) and along the $c$ axis ($M_c, H \parallel c$). (b) The field derivatives $dM_{ab}/dH$ and $dM_c/dH$ versus~$H$ obtained from the data in~(a), more clearly revealing the metamagnetic transitions at 4.5~T and~4.6~T for $M_{ab}$ and~$M_{c}$, respectively.}
\label{fig:MH_EuPd2As2_VSM}
\end{figure}

Because $M$ at $H=5.5$~T is much smaller than the theoretical $M_{\rm sat}$, we measured $M$ up to the higher field $H = 13.8$~T as shown at $T = 2$~K in Fig.~\ref{fig:MH_EuPd2As2_VSM}(a). These $M(H)$ data demonstrate  that $M$ does not reach $M_{\rm sat}$ up to fields of 13.8~T, as expected from Eq.~(\ref{Eq:HcCalc}). At $T=2$~K and $H = 13.8$~T, we find $M_{ab} = 4.58\,\mu_{\rm B}$/Eu for $H \perp c$ and $M_{c} = 4.41\,\mu_{\rm B}$/Eu for $H \parallel c$ which are $\leq 65$\,\% of the theoretical $M_{\rm sat}$ value.  The $M(H)$ data for both field directions show metamagnetic transitions at $H\sim5$~T, confirming the data in Fig.~\ref{fig:MH_EuPd2As2_2K}. The derivatives $dM_{ab}/dH$ and $dM_c/dH$ versus~$H$ are shown in Fig.~\ref{fig:MH_EuPd2As2_VSM}(b) which reflect the weak metamagnetic transitions near 4.5~T and~4.6~T for $H_{ab}$ and~$H_{c}$, respectively.  Within MFT, a spin flop transition only occurs for a collinear AFM structure if the field is aligned parallel to the ordering axis.  The fact that metamagnetic transitions are observed for both $H\parallel c$ and $H\perp c$ supports our hypothesis above that the magnetic structure below $T_{\rm N2}$ is both noncollinear and noncoplanar.

The high-field slopes of $M_{ab}\propto H$ and $M_c\propto H$ in Fig.~\ref{fig:MH_EuPd2As2_VSM} obtained from proportional fits of $M$ versus~$H$ for the field range 6.0~T~$ \leq H \leq 13.8$~T are $3.35\times10^{-1}~\mu_{\rm B}$/T\,Eu and $3.18\times10^{-1}~\mu_{\rm B}$/T\,Eu, respectively.  By extrapolating the proportional dependence of $M_{ab}(H)$ to the value $M_{\rm sat} = 7\,\mu_{\rm B}$/Eu, one obtains the extrapolated value of the critical field as
\be
H_{ab}^{\rm c} \approx\  21~{\rm T}
\label{Eq:HcExtrap}
\ee
for $H \perp  c$. This value is nearly the same as the above value of $H_{ab}^{\rm c}$ in Eq.~(\ref{Eq:HcCalc}) estimated from $\chi_{ab}(T_{\rm N})$ using MFT\@.

\begin{figure}
\includegraphics[width=3in]{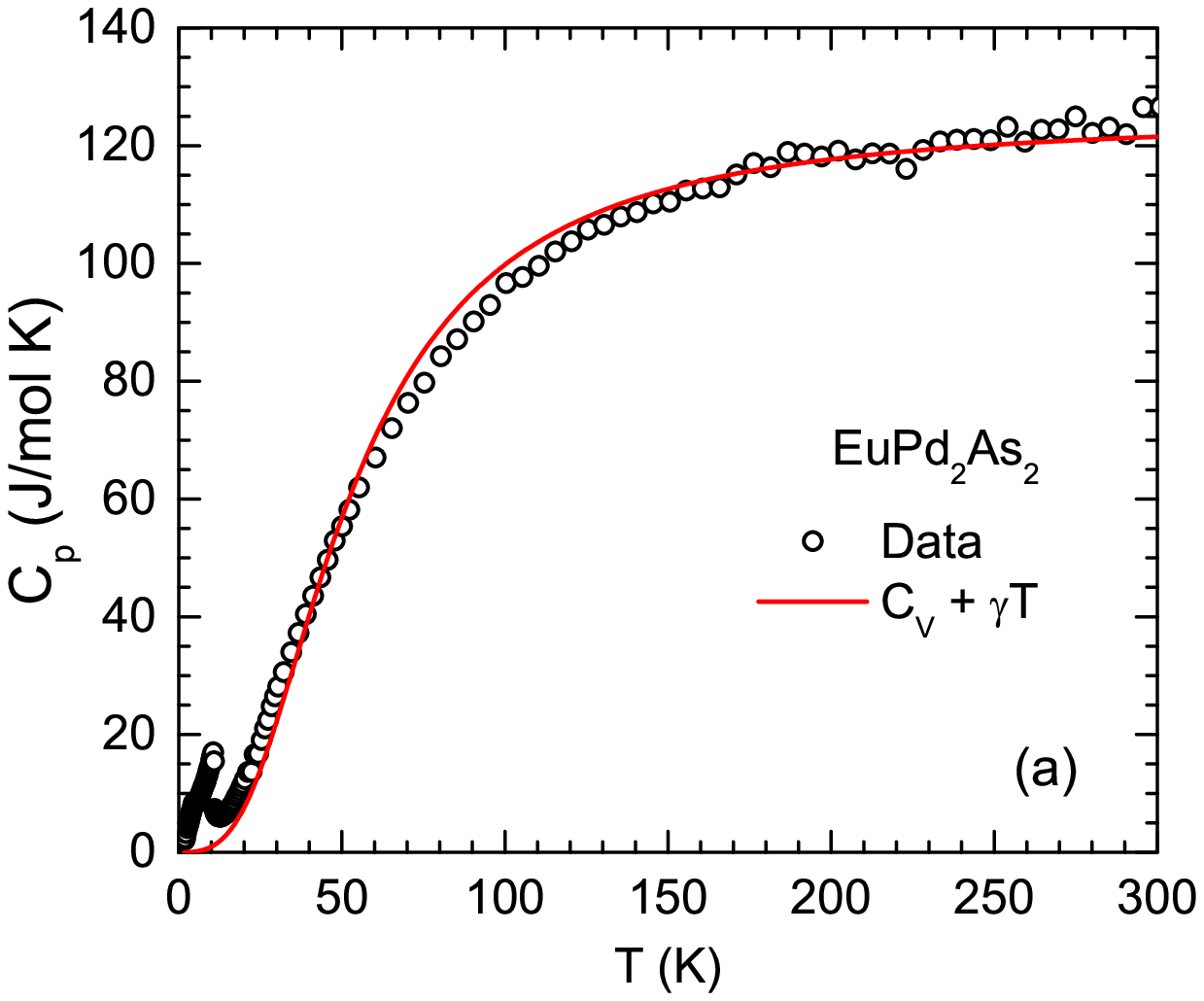}\vspace{0.1in}\vspace{-0.1in}
\includegraphics[width=3in]{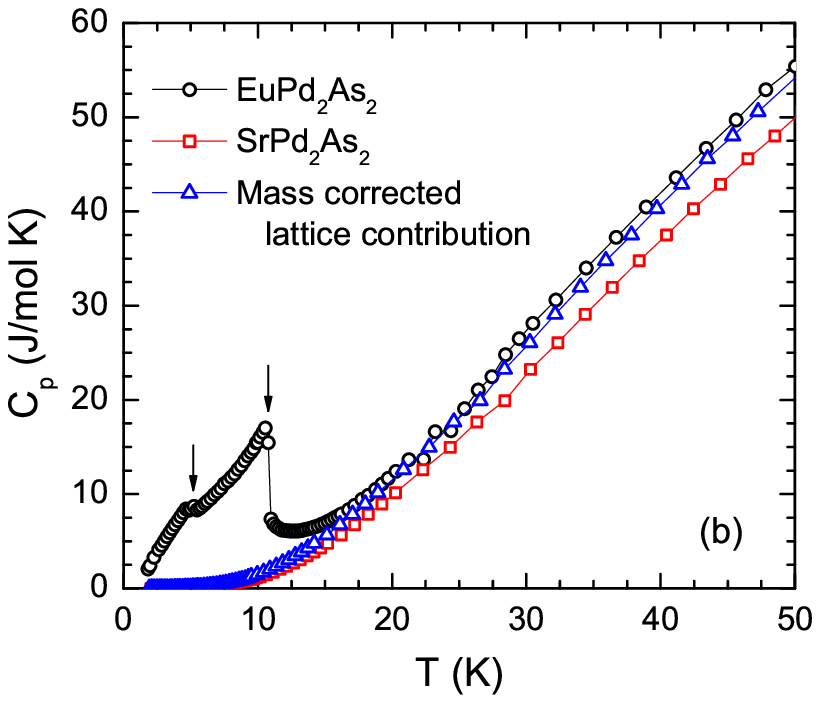}\vspace{0.1in}\vspace{-0.1in}
\includegraphics[width=3in]{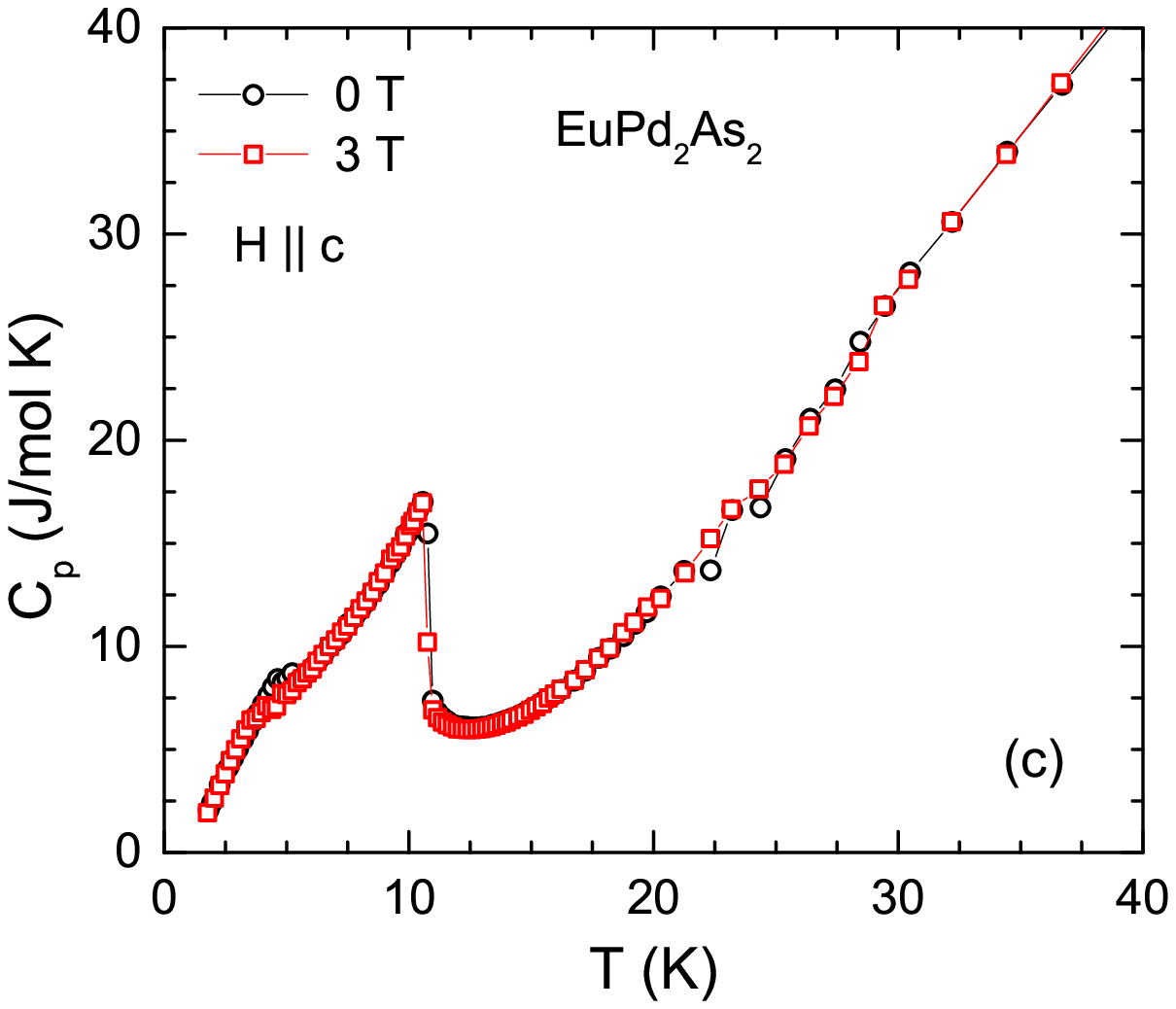}
\caption{(Color online) (a) Heat capacity $C_{\rm p}$ of a ${\rm EuPd_2As_2}$ crystal versus temperature $T$ from~1.8 to~300~K measured in $H=0$. The solid curve is a fit of the data from 16 to 300~K by the Debye lattice heat capacity $C_{\rm V\,Debye}(T)$ in Eq.~(\ref{eq:Debye_HC}). (b)~Expanded view of low-$T$ $C_{\rm p}(T)$ data in the temperature range 1.8~K~$\leq T\leq 50$~K. The $C_{\rm p}(T)$ data of ${\rm SrPd_2As_2}$ (Ref.~\cite{Anand2013a}) and the lattice contribution to $C_{\rm p}$ of ${\rm EuPd_2As_2}$ after correcting for the difference in formula weights of ${\rm EuPd_2As_2}$ and ${\rm SrPd_2As_2}$ are also shown. (c) Comparison of $C_{\rm p}(T)$ in magnetic fields $H=0$ and 3.0~T applied along the $c$ axis.}
\label{fig:HC_EuPd2As2}
\end{figure}

\subsection{\label{Sec:EuPd2As2_HC}Heat Capacity}

An overview of the $C_{\rm p}(T)$ data of an ${\rm EuPd_2As_2}$ crystal are shown in Fig.~\ref{fig:HC_EuPd2As2}(a). The low-$T$ $C_{\rm p}(T)$ data obtained in $H=0$ are shown on an expanded scale in Fig.~\ref{fig:HC_EuPd2As2}(b) and exhibit two clear anomalies near 5.5~K and 11~K, confirming the intrinsic nature of the AFM transitions at $T_{\rm N1}$ and $T_{\rm N2}$ revealed in the above $\chi(T)$ data.  The $C_{\rm p}(T)$ data measured at $H = 3.0$~T ($H \parallel c$) are compared with the data for $H=0$ in Fig.~\ref{fig:HC_EuPd2As2}(c). While no noticeable change is observed at $T_{\rm N1}$ between the $C_{\rm p}(T)$ at these two fields, the $T_{\rm N2}$ anomaly appears to broaden slightly with increasing field and to decrease slightly in  temperature at $H=3$~T compared to the zero-field data.  The weak field dependence in this field range is expected due to the much larger calculated and extrapolated critical fields in Eqs.~(\ref{Eq:HcCalc}) and~(\ref{Eq:HcExtrap}), respectively.

The zero-field $C_{\rm p}(T = 300~{\rm K})\approx 123$~J/mol\,K is close to the expected classical Dulong-Petit value $C_{\rm V} = 3nR = 15R = 124.7$~J/mol\,K at constant volume, \cite{Kittel2005, Gopal1966} where $n = 5$ is the number of atoms per formula unit (f.u.) and $R$ is the molar gas constant.  The $C_{\rm p}(T)$ data in the paramagnetic regime from 16 to 300~K were initially fitted by
\begin{equation}
C_{\rm p}(T) = \gamma T + n C_{\rm{V\,Debye}}(T),
\label{eq:Debye_HC-fit}
\end{equation}
where $\gamma T$ represents the electronic contribution to the heat capacity and $C_{\rm{V\,Debye}}(T)$ represents the Debye lattice heat capacity due to acoustic phonons at constant volume given by \cite{Gopal1966}
\begin{equation}
C_{\rm{V\,Debye}}(T) = 9 R \left( \frac{T}{\Theta_{\rm{D}}} \right)^3 {\int_0^{\Theta_{\rm{D}}/T} \frac{x^4 e^x}{(e^x-1)^2}\,dx}.
\label{eq:Debye_HC}
\end{equation}
Here we used the recently developed analytic Pad\'e approximant fitting function for $C_{\rm{V\,Debye}}(T)$. \cite{Goetsch2012} While fitting the data we first set $\gamma$ as an adjustable parameter which yielded $\gamma=2(3)$~mJ/mol\,K$^2$, so in the final fit we fixed $\gamma=0$. Thus in the final fit, the $C_{\rm p}(T)$  data were fitted with only one adjustable parameter $\Theta_{\rm D}$. The fit with 16~K~$\leq T \leq 300$~K gives $\Theta_{\rm D} = 216(2)$~K\@.  From a comparison of the data and the fit in Fig.~\ref{fig:HC_EuPd2As2}(a) shown by the solid red curve, the $C_{\rm p}(T)$ data in the paramagnetic state from 16~K up to 300~K are described reasonably well overall by the Debye model for the lattice heat capacity.

\begin{figure}
\includegraphics[width=3in]{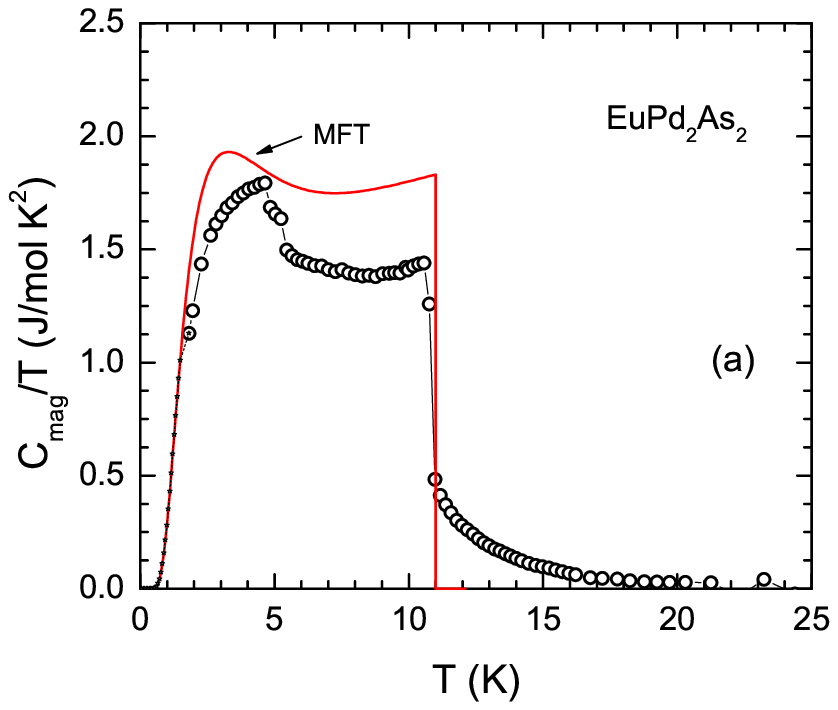}
\includegraphics[width=3in]{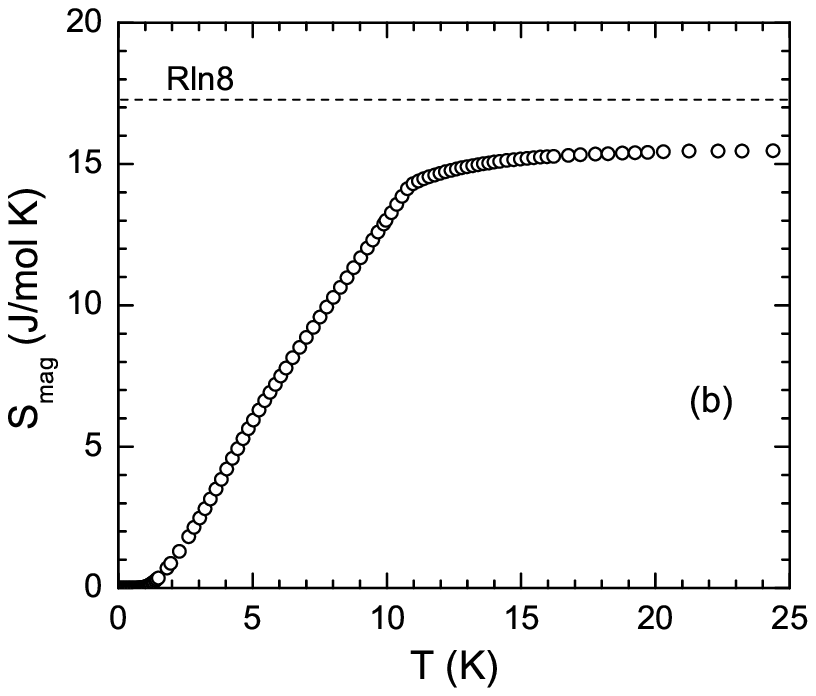}
\caption{(Color online) (a) Magnetic contribution to heat capacity $C_{\rm mag}$ (using mass corrected lattice contribution) for ${\rm EuPd_2As_2}$ plotted as $C_{\rm mag}(T)/T$ vs. $T$\@. The solid curve represents the mean-field theoretical value of $C_{\rm mag}$ for $S = 7/2$ and $T_{\rm N}= 11.0$~K\@ (b) Magnetic contribution to entropy $S_{\rm mag}(T)$.}
\label{fig:HC_EuPd2As2_mag}
\end{figure}

The magnetic contribution to the heat capacity $C_{\rm mag}(T)$ is estimated from the zero-field $C_{\rm p}(T)$ data of ${\rm EuPd_2As_2}$ by subtracting the lattice contribution.  As the lattice contribution we used the $C_{\rm p}(T)$ data of isostructural nonmagnetic ${\rm SrPd_2As_2}$. \cite{Anand2013a} The difference in formula weights of ${\rm EuPd_2As_2}$ and ${\rm SrPd_2As_2}$ was taken into account to estimate the lattice contribution to the heat capacity of ${\rm EuPd_2As_2}$. Since the lattice heat capacity is a function of $T/\Theta_{\rm D}$ and $\Theta_{\rm D}$ depends on formula mass $M$ ($\Theta_{\rm D} \sim 1/M^{1/2}$), the mass-corrected lattice contribution can be obtained by changing the temperature scale of $C_{\rm p}(T)$ to $T^*$, where
\begin{equation}
T^* = \frac{T}{(M_{\rm EuPd_2As_2}/M_{\rm SrPd_2As_2})^{1/2}}.
\label{eq:HC-DebyeT-scaling}
\end{equation}
The mass-corrected lattice contribution for ${\rm EuPd_2As_2}$ is shown in Fig.~\ref{fig:HC_EuPd2As2}(b).

The $C_{\rm mag}(T)$ of ${\rm EuPd_2As_2}$ is obtained by subtracting the $C_{\rm p}(T^*)$ lattice contribution of ${\rm SrPd_2As_2}$ from the measured $C_{\rm p}(T)$ data of ${\rm EuPd_2As_2}$ as shown in the plot of $C_{\rm mag}(T)/T$ versus~$T$ in Fig.~\ref{fig:HC_EuPd2As2_mag}(a).  Clear anomalies in $C_{\rm mag}(T)/T$ near $T_{\rm N1}=11$~K and $T_{\rm N2}=5.5$~K are apparent.  The nonzero $C_{\rm mag}(T)/T$ at $T > T_{\rm N1}$ in Fig.~\ref{fig:HC_EuPd2As2_mag}(a) indicates the presence of short-range AFM correlations above $T_{\rm N1}$.  The MFT prediction of  $C_{\rm mag}(T)/T$ for spin $S = 7/2$ and $T_{\rm N}= 11.0$~K is shown as the solid red curve in Fig.~\ref{fig:HC_EuPd2As2_mag}(a). \cite{Johnston2011a}  The magnetic entropy is the area under a $C_{\rm mag}(T)/T$ versus~$T$ plot.  It is seen that the missing experimental magnetic entropy at $T_{\rm N1}$ compared with the MFT prediction is largely recovered at $T>T_{\rm N1}$ where AFM correlations in the paramagnetic state contribute to the change in magnetic entropy.

In order to estimate $S_{\rm mag}(T)$ for 0~$<T<1.8$~K which is below our measurement temperature range, we extrapolated the $C_{\rm mag}(T)/T$ data to $T=0$ in accordance with the MFT prediction as shown by the dotted curve in Fig.~\ref{fig:HC_EuPd2As2_mag}(a). The magnetic contribution to the entropy $S_{\rm mag}(T)$ below 25~K was then determined by integrating the $C_{\rm mag}(T)/T$ versus $T$ data in Fig.~\ref{fig:HC_EuPd2As2_mag}(a) according to
\be
S_{\rm mag}(T) = \int_0^T\frac{C_{\rm mag}(T^\prime)}{T^\prime}\,dT^\prime,
\ee
as shown in Fig.~\ref{fig:HC_EuPd2As2_mag}(b).   It is seen from Fig.~\ref{fig:HC_EuPd2As2_mag}(b) that $S_{\rm mag}$ attains a value of 14.7~J/mol\,K at $T_{\rm N1}$ which is 85\% of the expected high-$T$ limit $R\ln(2S+1)=R\ln8 = 17.3$~J/mol\,K for $S=7/2$. The estimated experimental high-$T$ limit of $S_{\rm mag}$ is 89\,\% of $R\ln8$. In view of the magnetization data which indicated that the Eu is in the Eu$^{+2}$ oxidation state with $S = 7/2$ to high accuracy, the reduced value of $S_{\rm mag}$ compared with $R\ln(8)$ likely results from an inaccurate estimate of the lattice contribution used to obtain $C_{\rm mag}(T)$ from the measured $C_{\rm p}(T)$ data.

\subsection{\label{Sec:EuPd2As2_Rho} Electrical Resistivity}

\begin{figure}
\includegraphics[width=3in]{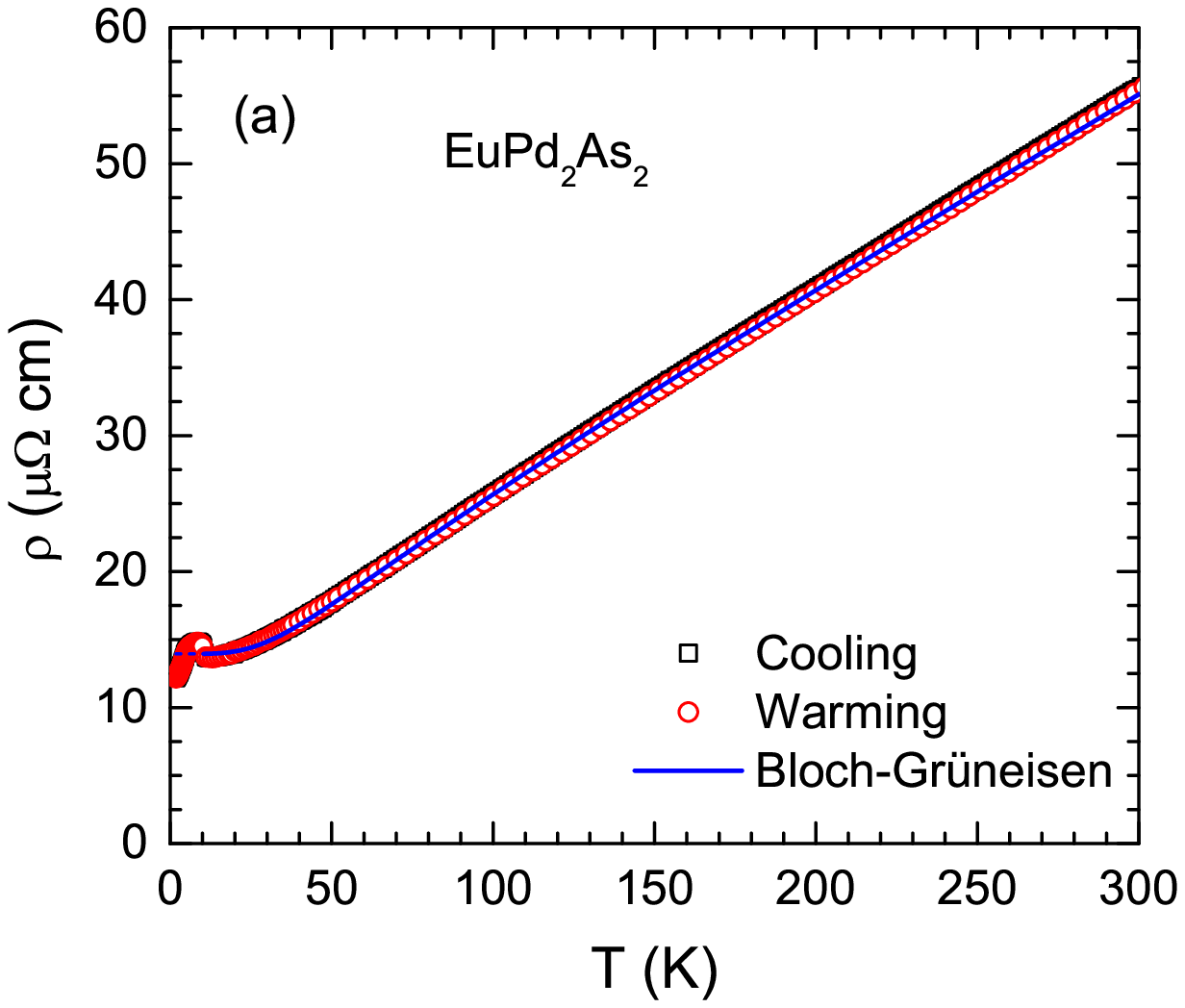}\vspace{0.1in}
\includegraphics[width=3in]{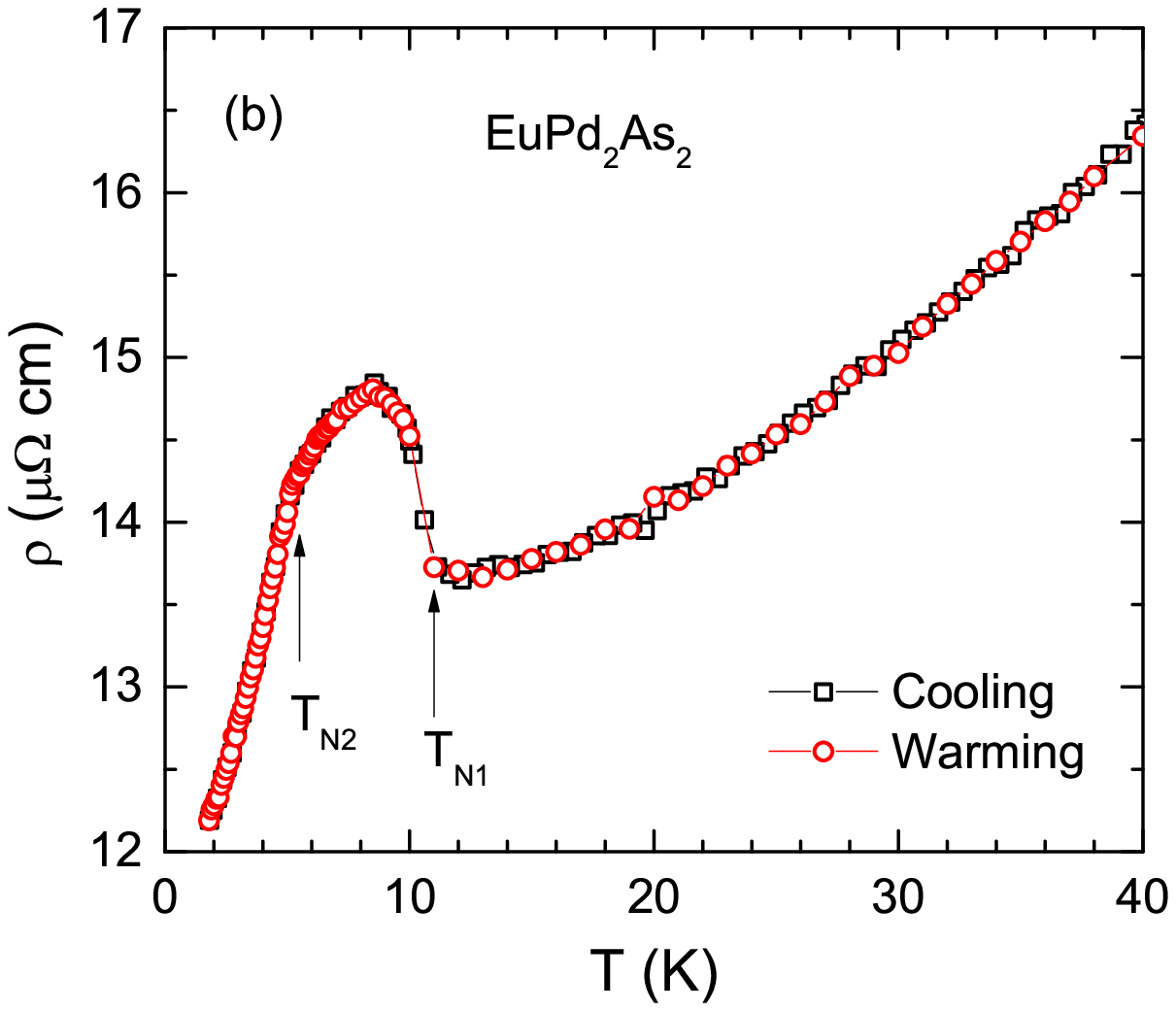}
\caption{(Color online) (a) In-plane electrical resistivity $\rho$ of a ${\rm EuPd_2As_2}$ single crystal in zero magnetic field versus temperature $T$ in the temperature range 1.8--300~K\@.  The solid blue curve is the fit of $\rho(T)$ by the Bloch-Gr\"{u}neisen model in Eqs.~(\ref{eq:BG})--(\ref{eq:BG_rhon}) for 12~K~$\leq T \leq$~300~K\@. The dashed curve is the extrapolation of the fit to $T =0$. (b) Expanded view of low-$T$ $\rho(T)$ data. }
\label{fig:rho_EuPd2As2}
\end{figure}

The $ab$-plane $\rho(T)$ data of a ${\rm EuPd_2As_2}$ crystal measured in zero magnetic field are shown in Fig.~\ref{fig:rho_EuPd2As2}. The low value of residual resistivity $\rho_0 = 12.2~\mu \Omega\,{\rm cm}$ at $T= 1.8$~K and the value of residual resistivity ratio ${\rm RRR} \equiv \rho(300\,{\rm K})/\rho(1.8\,{\rm K})\approx 4.5$ indicate a good quality of our single crystals. Metallic behavior is indicated from both the magnitude and $T$~dependence of $\rho$.

We fitted our paramagnetic-state zero-field $\rho(T)$ data by the Bloch-Gr\"uneisen (BG) model.  The BG resistivity $\rho_{\rm BG}$ due to the scattering of conduction electrons by acoustic lattice vibration is given by \cite{Blatt1968}
\label{Eqs:Allrho}
\begin{equation}
\rho_{\rm BG}(T/\Theta_{\rm R})= 4 \mathcal{R} \left( \frac{T}{\Theta _{\rm{R}}}\right)^5 \int_0^{\Theta_{\rm{R}}/T}{\frac{x^5}{(e^x-1)(1-e^{-x})}dx},
 \label{eq:BG}
\end{equation}
where $\mathcal{R}$ is a material-dependent prefactor and $\Theta_{\rm R}$ is the Debye temperature determined from resistivity data. One obtains
\begin{equation}
\rho_{\rm BG}(T/\Theta_{\rm R} = 1) = 0.9\,464\,635\,{\cal R}.
\label{eq:BG_R}
\end{equation}
The experimental $\rho(T)$ data were fitted by
\begin{equation}
\rho(T) = \rho_1 + \rho(\Theta_{\rm R}) \rho_{\rm n}(T/\Theta_{\rm R}),
\label{eq:BG_fit}
\end{equation}
where $\rho_1 = \rho_0 + \rho_{\rm sd}$ is the sum of $\rho_0$ and the spin-disorder resistivity $\rho_{\rm sd}$ due to the presence of disordered magnetic moments, and the normalized dimensionless BG resistivity $\rho_{\rm n}(T/\Theta_{\rm R})$ can be obtained from Eqs.~(\ref{eq:BG}) and~(\ref{eq:BG_R}) as
\begin{eqnarray}
\rho_{\rm n}(T/\Theta_{\rm R}) & =& 4.226\,259 \left( \frac{T}{\Theta _{\rm{R}}}\right)^5 \label{eq:BG_rhon}\\
 & & \times \int_0^{\Theta_{\rm{R}}/T}{\frac{x^5}{(e^x-1)(1-e^{-x})}dx}.\nonumber
\end{eqnarray}

We fitted the $\rho(T)$ data by Eqs.~(\ref{eq:BG_fit}) and~(\ref{eq:BG_rhon}) using the three independent fitting  parameters $\rho_1$, $\rho(\Theta_{\rm R})$ and $\Theta_{\rm R}$ for 12~K~$\leq T \leq$~300~K where we used the analytic Pad\'e approximant fitting function from Ref.~\onlinecite{Goetsch2012} for $\rho_{\rm n}(T/\Theta_{\rm R})$ in Eq.~(\ref{eq:BG_rhon}). A good fit of the $\rho(T)$ data was obtained with the fitting parameters $\rho_1 = 13.95(5)~\mu \Omega$\,cm, $\rho(\Theta_{\rm{R}}) = 24.1(4)~\mu \Omega$\,cm, and $\Theta_{\rm{R}} = 182(3)$~K, as shown by the solid blue curve in Fig.~\ref{fig:rho_EuPd2As2}(a).  The value $\mathcal{R} = 25.5~\mu \Omega$\,cm is obtained from the value of $\rho(\Theta_{\rm{R}})$ using Eq.~(\ref{eq:BG_R}) and $\rho_{\rm sd} \approx 1.8~\mu \Omega$\,cm is obtained from the value of $\rho_1$ using value $\rho_0(T = 1.8~{\rm K}) = 12.2~\mu \Omega$\,cm.  The value $\Theta_{\rm{R}} = 182(3)$~K is somewhat smaller than $\Theta_{\rm D} = 216(2)$~K obtained from the above analysis of the heat capacity data in the paramagnetic state in terms of the Debye model.  The values of $\Theta_{\rm{R}}$ and $\Theta_{\rm{D}}$ are not expected to be identical because of the different assumptions and approximations made in the Debye model of the lattice heat capacity and the Bloch-Gr\"uneisen model of the resistivity as outlined in Refs.~\onlinecite{Gopal1966, Goetsch2012, Blatt1968}.

From Fig.~\ref{fig:rho_EuPd2As2}(b), $\rho$ decreases with decreasing temperature in the paramagnetic state at $T > T_{\rm N1}$ but then sharply increases at $T = T_{\rm N1}$, reaches a maximum at $T = 9.0$~K and again starts decreasing with decreasing $T$ for $T < 9.0$~K with a rapid decrease below $T_{\rm N2}$.  We define the quantity $\Delta\rho$ to be the difference in $\rho$ between its value at the maximum of the peak and the value at~$T_{\rm N1}$.  Below about 4~K the resistivity is lower than the value extrapolated from above $T_{\rm N1}$.  No thermal hysteresis is observed between the heating and cooling cycles of $\rho$ measurements.  The increase in $\rho$ on decreasing $T$ below an AFM transition temperature has been observed in many systems and is usually attributed to the formation of superzone energy gaps within the Brillouin zone. \cite{Takabatake1998, Budko2000,Mun2010,Das2012,Pandey2009, Elliott1963, Elliott1964, Ellerby1998}

\begin{figure}
\includegraphics[width=3in]{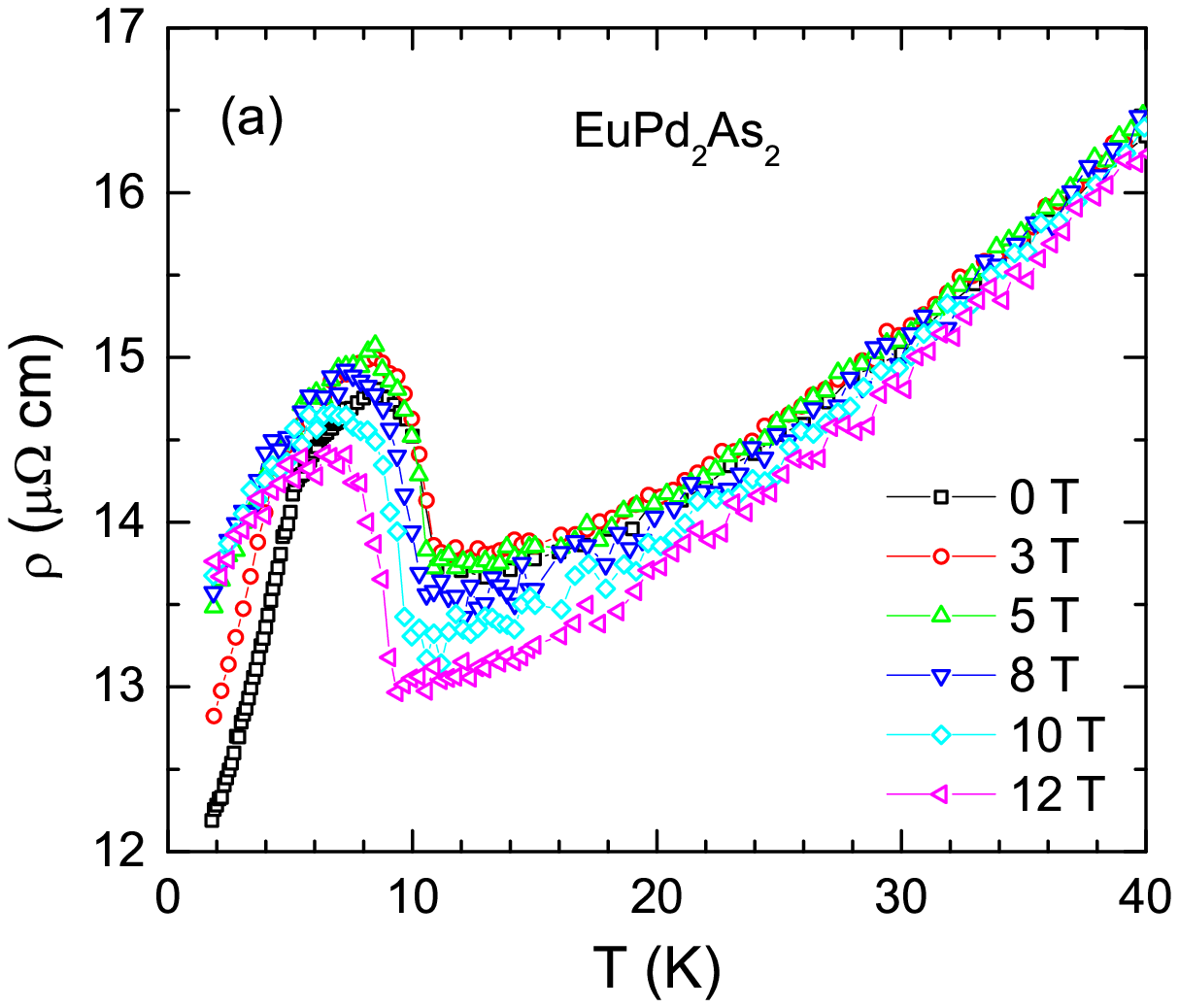}\vspace{0.1in}
\includegraphics[width=3in]{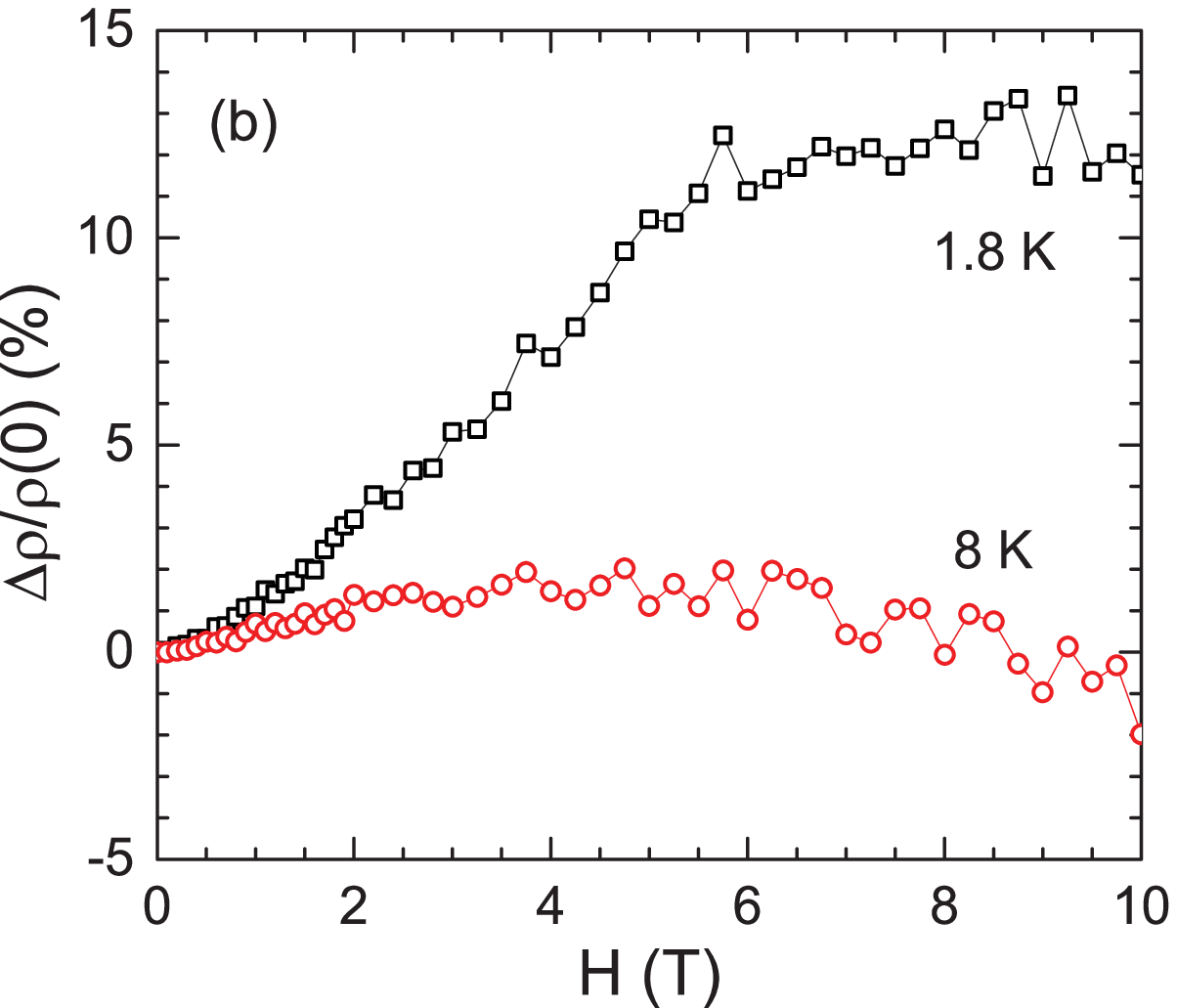}
\caption{(Color online) (a) $ab$-plane electrical resistivity $\rho$ of a ${\rm EuPd_2As_2}$ single crystal versus temperature $T$ in different magnetic fields~$H$ as indicated. (b) Magnetoresistance $\Delta\rho/\rho(0)\equiv [\rho(H)-\rho(0)]/\rho(0)$ versus $H$.}
\label{fig:rho_H_EuPd2As2}
\end{figure}

The $\rho(T)$ data measured at different $H$ are shown in Fig.~\ref{fig:rho_H_EuPd2As2}(a). We do not see any significant effect of magnetic field on $\Delta\rho$ below $T_{\rm N1}$ even at $H=12$~T\@. However, this field is still much smaller than the critical field estimated in Eqs.~(\ref{Eq:HcCalc}) and~(\ref{Eq:HcExtrap}).  As expected for an AFM system, $T_{\rm N1}$ decreases with increasing $H$ and in the paramagnetic state $\rho$ in the vicinity of $T_{\rm N1}$ decreases with increasing $H$ which shows a negative magnetoresistance (MR) behavior. For $T_{\rm N2} < T < T_{\rm N1}$ initially the $\rho$ increases weakly and then decreases with increasing $H$, although the change $\Delta \rho$ upon entering the antiferromagnetic state remains nearly unchanged. Thus there is no signature of suppression of the effect of magnetic superzone formation up to the maximum measurement field of 12~T\@. For $T < T_{\rm N2}$ the $\rho$ increases with increasing $H$ and thus a positive MR is observed.

The $H$ dependence of $\rho$ is shown in Fig.~\ref{fig:rho_H_EuPd2As2}(b) for 1.8~K and 8~K\@. The $\rho$ data are normalized as $\Delta\rho(H)/\rho(0) = [\rho(H)- \rho(0)]/\rho(0)$ to show the magnetoresistance behavior. The MR data are noisy but the basic trend of data can be inferred. At 1.8~K, initially the MR increases with increasing $H$ up to $\approx 5.5$~T above which the rate of increase decreases and eventually MR approaches a  constant value. The MR is positive throughout and is $\approx 12$\% at 10~T at 1.8~K\@. At 8~K the MR is weakly positive for $H \leq 8$~T above which it becomes negative as was also inferred from the $\rho(T)$ data measured at different $H$ shown in Fig.~\ref{fig:rho_H_EuPd2As2}(a). Because of the noise in the data it is not possible to determine the precise field at which this crossover from positive MR to negative MR takes place.

Usually superzone boundaries collapse with the application of a magnetic field and the effect of a superzone energy gap is suppressed. Contrary to this expectation, in the present compound the effect of the superzone energy gap persists up to the maximum investigated field of 12~T without any sign of a collapse of the superzone boundaries. A similar insensitiveness of the superzone gap to an external field has been observed in GdPd$_3$B$_{0.5}$C$_{0.5}$ where no change in the resistivity upturn behavior was noticed at 7~T. \cite{Pandey2009} In the case of GdPd$_3$B$_{0.5}$C$_{0.5}$ it was argued that the strength of the magnetic coupling between the moments is strong enough to prevent an effect of the external field. A similar situation may hold for the present compound because our maximum measurement field is roughly a factor of two smaller than the critical field.

\section{\label{Conclusion} Summary and Conclusions}

The physical properties of ${\rm EuPd_2As_2}$ single crystals were investigated using $\chi(T)$, $M(H,T)$, $C_{\rm p}(H,T)$ and $\rho(H,T)$ measurements.  The $\rho(T)$ data indicate metallic behavior.  The high-$T$ $\chi(T)$ data follow the Curie-Weiss law with a Curie constant consistent with Eu$^{+2}$ spins $S = 7/2$ with $g=2$ and Weiss temperature $\theta_{\rm p}\approx -30$~K indicative of dominant AFM interactions.   The $C_{\rm p}(T)$ data from 16 to~300~K are fitted well by the Debye theory of lattice heat capacity, yielding a Debye temperature $\Theta_{\rm D} = 216(2)$~K\@.  The $\rho(T)$ data from 12 to 300~K agree with the Bloch-Gr\"uneisen model of the resisitivity arising from electron-phonon scattering, where the fitted Debye temperature is $\Theta_{\rm R} = 182(3)$~K, somewhat smaller than the value obtained from analyzing the $C_{\rm p}(T)$ data.

At lower~$T$, the $\chi(T)$ data indicate long-range AFM ordering at $T_{\rm N1} = 11.0$~K with another transition at $T_{\rm N2} = 5.5$~K that is likely a spin reorientation transition.  The anisotropic $\chi(T)$ data for $T_{\rm N2}< T < T_{\rm N1}$ suggest a planar noncollinear AFM structure with the ordered moments aligned within the $ab$~plane, consistent with a helical or cycloidal magnetic structure with a turn angle of $\sim104^\circ$ or $\sim139^\circ$ between adjacent layers of ferromagnetically-aligned spins.  The anisotropic $\chi(T)$ and $M(H)$ isotherm data suggest that the AFM structure at $T < T_{\rm N2}$ becomes noncoplanar, with equal numbers of spins canting in opposite directions out of the $ab$ plane, thus preserving an overall AFM structure.  The $M(H)$ isotherm measurements for $H\parallel c$ and $H\perp c$ up to $H=14$~T at $T=2$~K both show weak metamagnetic transitions at $H\sim 5$~T\@.  Two estimates indicate that the critical field at which all Eu spins become aligned with the field with increasing field at 2~K is $H^{\rm c}\approx 22$~T, which is about 60\% larger than our maximum measurement field of 14~T\@.

The $C_{\rm p}(T)$ and $\rho(T)$ measurements show anomalies at both $T_{\rm N1}$ and~$T_{\rm N2}$.  Although $\rho$ decreases monotonically on cooling from 300~K to 10~K, it increases with decreasing~$T$ below $T_{\rm N1}$, suggesting that part of the Fermi surface becomes gapped due to the AFM ordering, and then decreases again below 9.0~K\@.  The $\rho(T)$ shows a 12\% positive magnetoresistance at $T = 1.8$~K and $H=10$~T, but the size of the upturn below $T_{\rm N1}$ is not affected by fields up to 12~T\@.

\acknowledgments

The research at Ames Laboratory was supported by the U.S. Department of Energy, Office of Basic Energy Sciences, Division of Materials Sciences and Engineering.  Ames Laboratory is operated for the U.S. Department of Energy by Iowa State University under Contract No.~DE-AC02-07CH11358.

\end{document}